\documentclass{aa}
\usepackage{times}
\usepackage{xspace}
\usepackage{natbib}
\usepackage{graphicx}
\usepackage{amsmath}
\usepackage{amssymb}
\usepackage{url}

\newcommand{\efold}{\ensuremath{E_{\text{F}}}\xspace}
\newcommand{\batse}{\textsl{BATSE}\xspace}
\newcommand{\vela}{Vela X-1\xspace}
\newcommand{\ca}{\mbox{$\sim$}}
\newcommand{\kevnxs}{ke\kern -0.09em V}
\newcommand{\kev}{\kevnxs\xspace}
\newcommand{\err}[2]{\ensuremath{^{+#1}_{-#2}}\xspace}
\newcommand{\Rsun}{\ensuremath{\mbox{R}_\odot}\xspace}
\newcommand{\Msun}{\ensuremath{\mbox{M}_\odot}\xspace}
\newcommand{\hd}{HD\,77581\xspace}
\newcommand{\hexe}{\textsl{HEXE}\xspace}
\newcommand{\osse}{\textsl{OSSE}\xspace}
\newcommand{\ginga}{\textsl{Ginga}\xspace}
\newcommand{\sax}{\textsl{BeppoSAX}\xspace}
\newcommand{\xte}{\textsl{RXTE}\xspace}
\newcommand{\pca}{\textsl{PCA}\xspace}
\newcommand{\hexte}{\textsl{HEXTE}\xspace}
\newcommand{\asm}{\textsl{ASM}\xspace}
\newcommand{\qcm}{\ensuremath{\text{cm}^2}\xspace}
\newcommand{\nh}{\ensuremath{N_{\text{H}}}\xspace}
\newcommand{\ev}{e\kern -0.11em V\xspace}

\newcommand{\xmm}{\textsl{XMM-Newton}\xspace}

\begin{document}
\title{Confirmation of Two Cyclotron Lines in Vela X-1}

\author{Ingo Kreykenbohm\inst{1} \and Wayne Coburn\inst{2} \and J\"orn
  Wilms\inst{1} \and Peter Kretschmar\inst{4,5}\and R\"udiger
  Staubert\inst{1} \and William A. Heindl\inst{3} \and Richard E.
  Rothschild\inst{3}}
 
\offprints{kreyken@astro.uni-tuebingen.de}

\institute{Institut f\"ur Astronomie und Astrophysik -- Astronomie,
  Sand 1, D-72076 T\"ubingen, Germany 
  \and Space Sciences Laboratory,
  University of California, Berkeley, Berkeley, CA, 94702-7450, U.S.A.
  \and Center for Astrophysics and Space Sciences, University of
  California at San Diego, La Jolla, CA 92093-0424, U.S.A. \and INTEGRAL
  Science Data Center, 6 ch.\ d'\'Ecogia, CH-1290 Versoix, Switzerland
  \and Max-Planck-Institut f\"ur extraterrestrische Physik,
  Giessenbachstr.~1, D-85740 Garching, Germany}

\date{Received 3 August 2001 / Accepted }

\abstract{We present pulse phase-resolved X-ray spectra of the high
  mass X-ray binary \vela using the Rossi X-ray Timing Explorer.  We
  observed \vela in 1998 and 2000 with a total observation time of
  \ca90\,ksec. We find an absorption feature at
  23.3\err{1.3}{0.6}\,\kev in the main pulse, \emph{that we interpret
    as the fundamental cyclotron resonant scattering feature (CRSF)}.
  The feature is deepest in the rise of the main pulse where it has a
  width of 7.6\err{4.4}{2.2}\,\kev and an optical depth of
  0.33\err{0.06}{0.13}.  This CRSF is also clearly detected in the
  secondary pulse, but it is far less significant or undetected during
  the pulse minima. We conclude that the well known CRSF at
  50.9\err{0.6}{0.7}\,\kev, which is clearly visible even in
  phase-averaged spectra, is the first harmonic and \emph{not} the
  fundamental. Thus we infer a magnetic field strength of $B=2.6\times
  10^{12}$\,G.  \keywords{X-rays: stars -- stars: magnetic fields --
    stars: pulsars: individual: Vela X-1} }

\maketitle

\section{Introduction}
\label{intro}
\vela (4U\,0900$-$40) is an eclipsing high mass X-ray binary (HMXB)
consisting of the B0.5Ib supergiant HD\,77581 and a neutron star with
an orbital period of 8.964\,days \citep{kerkwijk95a} at a distance of
$\sim$2.0\,kpc \citep{nagase89a}. The optical companion has a mass of
$\sim$23\,\Msun and a radius of $\sim$30\,\Rsun \citep{kerkwijk95a}.
Due to the small separation of the binary system (orbital radius:
$1.7\,{\rm R_*}$), the 1.4\,\Msun neutron star \citep{stickland97a} is
deeply embedded in the intense stellar wind \citep[$4 \times
10^{-5}$\,\Msun$\text{yr}^{-1}$;][]{nagase86a} of \hd.

The neutron star has a spin period of $\sim$283\,s
\citep{rappaport75a,mcclintock76a}.  Both spin period and period
derivative have changed erratically since the first measurement as is
expected for a wind accreting system. The last measurements with the
Burst and Transient Source Experiment\footnote{See\hfill
  \url{http://www.batse.msfc.nasa.gov/batse/pulsar/data/sources/velax1.html}}
(\batse) resulted in a period of \ca283.5\,s.

The X-ray luminosity of \vela is \ca$4 \times 10^{36}\,{\rm erg\,
  s}^{-1}$, which is a typical value for an X-ray binary.  Several
observations have shown that the source is extremely variable 
with flux reductions to less than 10\,\% of its normal value
\citep{kreykenbohm99a,inoue84a}.  In these instances it is not known
whether a clump of material in the wind blocks the line of sight or if
the accretion itself is choked.

The phase averaged X-ray spectrum of Vela X-1 has been modeled with a
power law including an exponential cutoff \citep{white83a,tanaka86a}
or with the Negative Positive EXponential \citep[NPEX-model;][see also
Eq.~\ref{npex} below]{mihara95a}.  The spectrum is further modified by
strongly varying absorption which depends on the orbital phase of the
neutron star \citep{kreykenbohm99a,haberl90a}, an iron fluorescence
line at 6.4\,\kev, and occasionally an iron edge at 7.27\,\kev
\citep{nagase86a}. A cyclotron resonant scattering feature (CRSF) at
\ca55\,\kev was first reported from observations with \hexe
\citep{kendziorra92a}. \citet{makishima92a} and \citet{choi96a}
reported an absorption feature at \ca25\,\kev to 32\,\kev from \ginga
data.  \citet{kreykenbohm99a} using observations with the Rossi X-ray
Timing Explorer (\xte) and more detailed analysis of older \hexe and
\ginga data \citep{kretschmar96a,mihara95a} supported the existence of
both lines, while \citet{orlandini97c} reported only one line at
\ca55\,\kev based on \sax observations. Later observations with \xte
also cast some doubt on the existence of the 25\,\kev line
\citep{kreykenbohm00b}.

CRSFs are due to resonant scattering of electrons in Landau levels in
the strong magnetic field ($B \ca 10^{12}$\,G) of neutron stars
\citep[see][and references therein]{coburn01a,araya99a,meszaros85a}.
The energy of the CRSF is given by
\begin{equation}
E_{\rm C}=11.6\text{\,\kev} \times \frac{1}{1+z} \times \frac{B}{10^{12}\,\text{G}}
\end{equation}
$E_{\rm C}$ is the centroid energy of the CRSF in \kev, $z$ is the
gravitational redshift, and $B$ the magnetic field strength.
Therefore, knowing the correct fundamental energy of the CRSF is
important for inferring the correct magnetic field strength. This
helps in understanding the emission mechanisms of accretion powered
X-ray pulsars and the effects of the magnetic field
on the X-ray production.

In this paper, Sect.~\ref{data} describes the instruments on \xte,
the relevant calibration issues, the software used for the analysis,
and the data.  Sect.~\ref{spectra} gives an overview of the spectral
models used, introduces pulse phase resolved spectroscopy, and
discusses the evolution of the spectral parameters over the pulse in 16
phase bins. Later we use fewer bins to increase the significance and
discuss the behavior of the 25\,\kev and 50\,\kev CRSFs in different
pulse phases.  In Sect.~\ref{discussion} we discuss the existence of
the 25\,\kev line and review our findings in the context of the other
cyclotron sources.

\section{Data}
\label{data}
\subsection{Observations}
\begin{figure}
\includegraphics[width=\columnwidth]{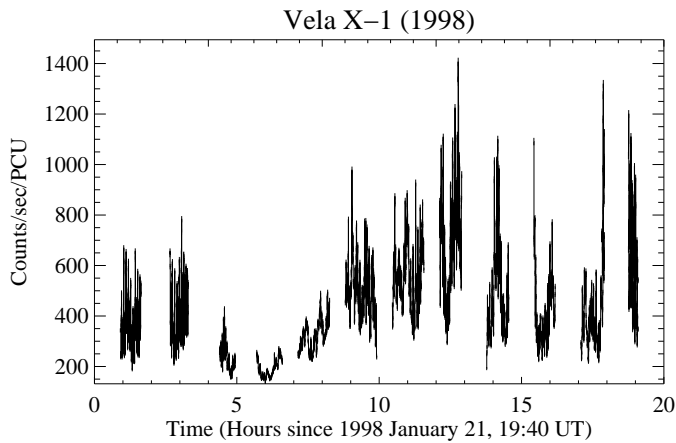}
\caption{\label{ao3light}RXTE PCA light-curve of the 1998 observation. 
  The variability is not only due to the flaring, but also due to the
  283\,s pulse.  In this and all following light-curves, the count rate
  has been normalized to one PCU. The temporal resolution of the
  light-curves is 16\,s and covers the entire energy range of the
  \pca. The gaps in the light-curve are due to Earth occultation and
  passages through the South Atlantic Anomaly. Note the extended low
  in the first half of the observation: it is probably due to a
  massive blob of cold material passing through the line of sight. As
  this paper is dedicated to phase-resolved spectroscopy and the
  existence of the CRSFs, this data was excluded from our analysis and
  we do not discuss this phenomenon here. For a detailed discussion,
  see \citet{kretschmar99a}. After the extended low, \vela was flaring
  and remained so until the end of the observation.}
\end{figure}

\begin{figure}
\includegraphics[width=\columnwidth]{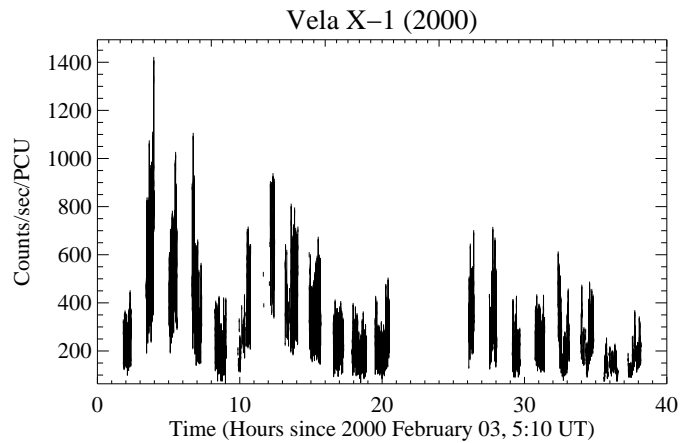}
\caption{\label{ao4light}RXTE PCA light-curve of the 2000 observation
  as for the 1998 observation in Fig.~\ref{ao3light}. Gaps in the 
  light-curve are due to SAA, Earth occultation, and \xte observing
  other sources.}
\end{figure}

We observed \vela with the \xte in 1998 and again in 2000. The first
observation in \xte-AO3 was made 1998 January 21/22 (JD\,2450835.32 --
JD\,2450836.05) resulting in 30\,ksec on-source time. During this
observation we encountered an extended low (see Fig.~\ref{ao3light});
therefore we only used data taken after the extended low, starting
about 8~hours after the beginning of the observation at JD\,2450835.64
(see Fig.~\ref{ao3light}).  The second observation was in \xte-AO4 and took
place 2000 February 03/04 (JD 2451577.71 -- JD 2451579.22) resulting
in 60\,ksec on-source time.  Figs.~\ref{ao3light} and~\ref{ao4light}
show the \pca light-curves of the 1998 and 2000 observations.

\subsection{The Instruments: \xte}

\xte \citep{bradt93a} consists of three instruments: the Proportional
Counter Array \citep[\pca;][]{jahoda96a}, the High Energy X-ray Timing
Experiment \citep[\hexte;][]{rothschild98a}, and the All Sky Monitor
\citep[\asm;][]{levine96a}. 

\begin{figure}
\includegraphics[width=\columnwidth]{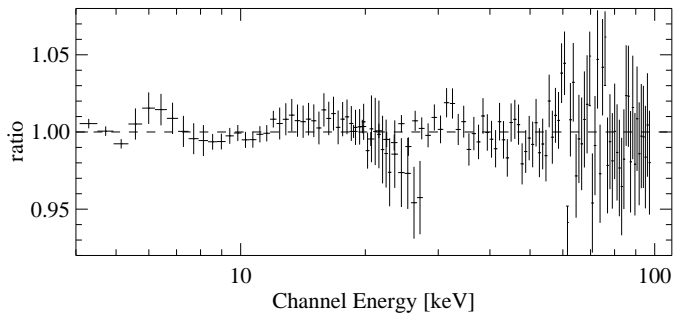}
\caption{\label{crabsys}Ratio of PCA and HEXTE data to the best fit
model for a Crab observation (obsid 40805-01-05-01,
1999 December~12).  The two power laws, which represent the nebula and
pulsar, had photon indices $\Gamma_1 = 2.25\err{0.03}{0.03}$ and
$\Gamma_2 = 1.88\err{0.05}{0.03}$ respectively.  A single power law model gave
$\Gamma=2.13\pm0.02$, which is in agreement with recent observations
by \xmm \citep{willingale01a}.  The \pca response matrix was created
with \textsl{PCARSP 2.43}.
}
\end{figure}

The \pca consists of five co-aligned Xenon proportional counter units
(PCUs). It has a total effective area of \ca6000\,\qcm and a nominal
energy range from 2 to 60\,\kev. Background subtraction in the \pca is
done by a model as the instrument is constantly pointed at the source.
We used the \textsl{Sky\_VLE}-model \citep[see][for a
description]{stark97b}, as is appropriate for the count rate of \vela.
The response matrix was generated with \textsl{PCARSP}, version 2.43.
To account for the uncertainties in the \pca-response matrix, we used
systematic errors as given in Table~\ref{syserror}. We derived these
numbers by fitting a two power law model simultaneously to a \pca and
\hexte spectrum of the Crab nebula and pulsar (see Fig.\ref{crabsys}).
See \citet{wilms99b} for a discussion of this procedure.

After the analysis was completed and while this paper was in
preparation, \textsl{FTOOLS Patch~5.0.4} was released, introducing
\textsl{PCARSP} version~7.10.  Tests using the new response matrix,
however, show that the best-fit parameters are only marginally
different compared to the old matrices, with the only difference being
a \ca30\,\% larger value of the hydrogen column and virtually no
difference at higher energies.

\begin{table}
\caption{\label{syserror}Systematic errors applied to the \pca-data to
  account for the uncertainties in the 2.43 and 7.10 version of the
  \pca response matrices. We used the response matrix generated by
  \textsl{PCARSP~2.43} to derive the results presented here (see
  text).} 
\begin{tabular}{cccc}
\hline
Channels & \kev & Syst. (2.43) & Syst. (7.10)\\
\hline 
\phantom{1}0  --   \phantom{1}9 &  \phantom{1}1 --  \phantom{1}5 & 1.0\,\% & 0.3\,\% \\
10 --   15 &  \phantom{1}5 --  \phantom{1}8 & 1.0\,\% & 1.3\,\% \\
16 --   26 &  \phantom{1}8 -- 12 & 0.5\,\% & 1.3\,\% \\
27 --   39 & 12 -- 18 & 0.5\,\% & 0.3\,\% \\
40 --   58 & 19 -- 30 & 2.0\,\% & 2.0\,\% \\
\hline
\end{tabular}
\end{table}

Our light-curves and phase-resolved spectra were obtained from binned
mode data with a temporal resolution of 250\,ms and 128 spectral
channels. Due to the gain correction present in these data in the PCA
experiment data system, some pulse height analyzer (PHA) channels
remain empty. This problem is known and handled properly by
\textsl{PCARSP} (Tess Jaffe, priv.\ comm.). For clarity, these zero
counts/sec channels are not shown in our figures.  We therefore used
the \pca in the energy range from 3.5--21\,\kev (corresponding to PHA
channels 6--45) in our analysis and relied on the \hexte data above
this energy range.

The \hexte consists of two clusters of four NaI(Tl)/CsI(Na)-Phoswich
scintillation detectors each, which are sensitive from 15\,\kev to
250\,\kev. The two clusters have a total net detector area of
1600\,\qcm. See \citet{rothschild98a} for a detailed description of
the instrument.  Early in the mission an electronics failure left one
detector in the second cluster unusable for spectroscopy. Background
subtraction in \hexte is done by source-background swapping of the two
clusters every 32\,s throughout the observation \citep{gruber96a}.
\hexte was used in the standard modes with a time resolution of
8\,$\mu$s and 256 spectral channels.  The response matrices were
generated with \textsl{HXTRSP}, version 3.1.  We used \textsl{HXTDEAD}
Version 2.0.0 to correct for the dead time.  To improve the statistical
significance of our data, we added the data of both \hexte clusters
and created an appropriate response matrix by using a 1:0.75 weighting
to account for the loss of a detector in the second cluster. This can
be safely done as the two cluster responses are very similar. To
further improve the statistical significance of the \hexte data at
higher energies, we binned several channels together as given in
Table~\ref{binning}. We chose the binning as a compromise between
increased statistical significance while retaining a reasonable energy
resolution. The binning resulted in statistically significant data
from 17\,\kev up to almost 100\,\kev (we detect the source at
\ca75\,\kev with a 3\,$\sigma$ significance in the pulse phase
resolved spectra and 5\,$\sigma$ in the phase averaged spectra) in the
2000 observation.

\begin{table}
\caption{\label{binning}Binning of the \hexte data. We chose the binning to
  get good statistical significance for each bin while also maintaining
  good energy resolution. Therefore we did not bin the channels below
  40\,\kev. Factor is the number of channels binned together. 
	Raw channels are approximately 1\,\kev in width.}
\begin{tabular}{cr}
\hline
Channels & Factor \\
\hline
\phantom{11}1   -- \phantom{1}39 & 1 \\
\phantom{1}40  --  \phantom{1}69 & 3 \\
\phantom{1}70  -- 109 & 8 \\
110 -- 255 & 32 \\
\hline
\end{tabular}
\end{table}

All analysis was done using the \textsl{FTOOLS} version 5.0.1 dated
2000 May 12. We used \textsl{XSPEC} 11.0.1aj, \citep{arnaud96a} with
several custom models (see below, Sect.~\ref{models}) for the
spectral analysis of the data.  To create phase-resolved spectra we
modified the \textsl{FTOOL fasebin}, to take the \hexte dead-time into
account. A version of this modified tool is available by contacting
the authors. 

\section{Spectral Analysis}
\label{spectra}
\subsection{Spectral Models}
\label{models}

Since probably all accretion powered X-ray pulsars have locally
super-Eddington flux at the polar cap \citep{nagase89a}, a realistic,
self-consistent calculation of the emitted spectrum is extremely
difficult.  Both radiative transfer and radiation hydrodynamics have
to be taken into account at the same time \citep{isenberg97a}.
Although this problem has been investigated for many years, there
still exists no convincing theoretical model for the continuum of
accreting X-ray pulsars \citep[and references therein]{harding94a}.
Most probably, the formation of the overall spectral shape is
dominated by \emph{resonant} Compton scattering
\citep{nagel81a,meszaros85a,brainerd:91a,sturner94a,alexander96a}.
This should produce a roughly power-law continuum with an exponential
cutoff at an energy characteristic of the scattering electrons.
Deviations should appear at the cyclotron resonant energy, i.e.\ the
cyclotron lines. Because the process is resonant scattering, as
opposed to absorption, it is natural to call ``cyclotron lines''
``cyclotron resonant scattering features''.  Because of the
computational complexities associated with modeling the continuum and
CRSF formation, this type of Comptonization has been far less studied
than \emph{thermal} Comptonization \citep{sunyaev80a,hua:95a}, and
empirical models of the continuum continue to be the only option for
data analysis.

Due to the general shape of the continuum spectrum, the empirical
models all approach a power-law at low energies, and have some kind of
cutoff at higher energies \citep{white83a,tanaka86a}. The CRSFs are
either modeled as subtractive line features or through multiplying the
continuum with a weighting function at the cyclotron resonant energy.
As we have shown previously \citep{kretschmar96b}, a smooth transition
between the power law and the exponential cutoff is necessary to avoid
artificial, line-like features in the spectral fit. This transition
region is notoriously difficult to model.  See \citet{kreykenbohm99a}
for a discussion of spectral models with smooth ``high energy
cutoffs'', such as the Fermi-Dirac cutoff \citep{tanaka86a} used in
our analysis of the AO1 data \citep{kreykenbohm99a}, and the Negative
Positive Exponential model \citep[NPEX;][]{mihara95a,mihara98a}.

In this paper, we describe the continuum of Vela~X-1,
$I_{\text{cont}}$, using the NPEX model,
\begin{equation}
 I_{\text{cont}}(E)   \propto   \left( E^{-\Gamma_1} + \alpha
   E^{+\Gamma_2}\right) \times \text{e}^{-E/E_{\text{F}}} \label{npex}
\end{equation}
where $\Gamma_{1}>0$ and $\Gamma_2=2$ \citep{mihara95a} and where
$E_{\text{F}}$ is the folding energy of the high energy exponential
cutoff.  In our experience, this model is the most flexible of the
different continuum models. For \vela, the continuum
between 7 and 15\,\kev is better described with the NPEX
model than the Fermi-Dirac cutoff.  The CRSFs are taken into account
by a multiplicative factor
\begin{equation}
I(E)  \propto  I_{\text{cont}}\exp\left(-\tau_{\text{GABS}}(E)\right)
\end{equation}
where the optical depth, $\tau_{\text{GABS}}$, has
a Gaussian profile \citep[][Eqs.~6,7]{coburn02a},
\begin{equation}\label{gabs}
\tau_{\text{GABS}}(E) = \tau_{\text{C}}
   \times  \exp\left(
 -\frac{1}{2}\left(\frac{E-E_{\text{C}}}{\sigma_{\text{C}}}\right)^2
                \right)
\end{equation}
Here $E_{\text{C}}$ is the energy, $\sigma_{\text{C}}$ the width, and
$\tau_{\text{C}}$ the optical depth of the CRSF.

We note that the relationship between the parameters of the empirical
continuum models and real physical parameters is not known. Because
thermal Comptonization spectra get harder with an increasing
Compton-$y$-parameter, it seems reasonable to relate the
characteristic frequency of the exponential cutoff, $E_{\text{F}}$, with
some kind of combination between the (Thomson) optical depth and the
temperature \citep[see, e.g.,][]{mihara98a}.  Because of the
complexity of the continuum and line formation, however, we caution
against a literal interpretation of any of the free parameters that
enter these continuum models, and will not attempt to present any
interpretation here (note, however, that the cyclotron line parameters
do have a physical interpretation).

\subsection{Pulse Phase-resolved spectra}

During the accretion process, material couples to the strong magnetic
field of the neutron star at the magnetospheric radius. This material
is channeled onto the magnetic poles of the neutron star, where spots
on the surface of the neutron star are created \citep[][and references
therein]{burnard91a}. If the magnetic axis is offset from the spin
axis, the rotation of the neutron star gives rise to pulsations
\citep{davidson73a}, as a distant observer sees the emitting region
from a constantly changing viewing angle.  Because the emission is
highly aniostropic, the spectra observed during different pulse phases
vary dramatically.  This is especially true for CRSFs whose
characteristics depend strongly on the viewing angle with respect to
the magnetic field \citep[for an in-depth discussion see][and
references therein]{araya00a,araya99a,isenberg97a}.  This dependence
can in principle be used to infer the geometric properties of the
emission region and the electron temperature.

To obtain phase-resolved spectra, the photon or bin times were
bary-centered and corrected for the orbital motion of Vela~X-1 using
the ephemeris of \citet{nagase89a}. Then the pulse phase was
calculated for each time bin (for the \pca) or for each photon (for
the \hexte) and then attributed to the appropriate phase bin.  Using
epoch folding \citep{leahy83a}, we determined the pulse period for
each observation individually. We derived pulse periods of
$P_{\text{pulse}}=283.5\pm0.1$\,s for the 1998 data and
$P_{\text{pulse}}=283.2\pm0.1$\,s for the 2000 data (using
$\text{JD} 2444279.546637$ as the zero phase for both observations).

\begin{figure}
\includegraphics[width=\columnwidth]{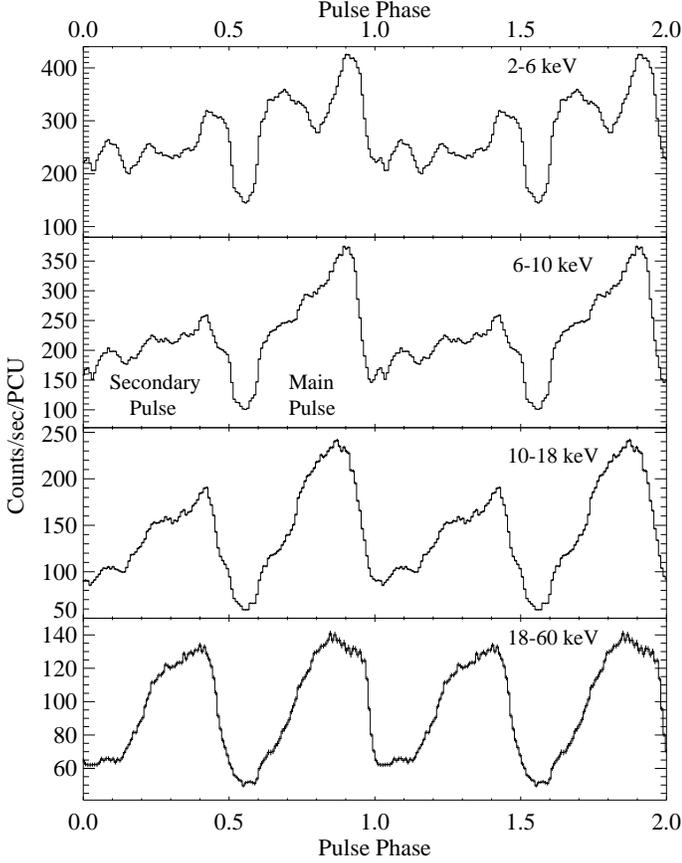}
\caption{\label{profileao4}Energy resolved pulse profiles in four
  bands derived from the 2000 data. At lower energies (below 6\,\kev),
  the pulse profile is very complex and shows a five peaked structure:
  two peaks in the main pulse and three in the secondary pulse.
  Between 6\,\kev and 10\,\kev, the first peak of the main pulse is
  reduced to a mere shoulder, while the second peak is very strong.
  Above 10\,\kev the pulse profile evolves continuously into a simple
  double pulse.  }
\end{figure}

Since the first detection of pulsations from \vela by
\citet{mcclintock76a}, the pulse profile is known to be very complex.
Above 20\,\kev, the pulse profile consists of two peaks which are
about equally strong. Since \vela has a relatively hard spectrum,
\citet{staubert80a} were able to detect the two pulses up to 80\,\kev.
We designate these two pulses as the \emph{main pulse} and the
\emph{secondary pulse} (see Fig.~\ref{profileao4} for the definition).

Below 5\,kev, these two pulses evolve into five peaks: the main pulse
develops into two peaks while the secondary pulse develops into three
peaks. Although the two pulses are almost equally strong at high
energies, the second peak of the main pulse is almost a factor of two
brighter at lower energies than the first two peaks of the secondary
pulse, while the third peak of the secondary pulse is almost as strong
as the first peak of the main pulse. For a detailed description of the
pulse profile and the relative intensities of the different peaks, see
\citet{raubenheimer90a} and Sect.~\ref{discussion_profile}.

In the following we first study the evolution of the continuum
parameters (i.e., the NPEX-model and the photo-electric absorption) and
then concentrate on the behavior of the CRSFs in the 2000 and 1998
observations using fewer phase bins.

\subsubsection{Spectral Evolution over the Pulse}

We used 16 phase bins for the 1998 as well as the 2000 data. The better
statistics of the 2000 observation would allow finer resolution (e.g.,
32 phase bins) but would make the comparison of the observations more
difficult.

\begin{figure}
\includegraphics[width=\columnwidth]{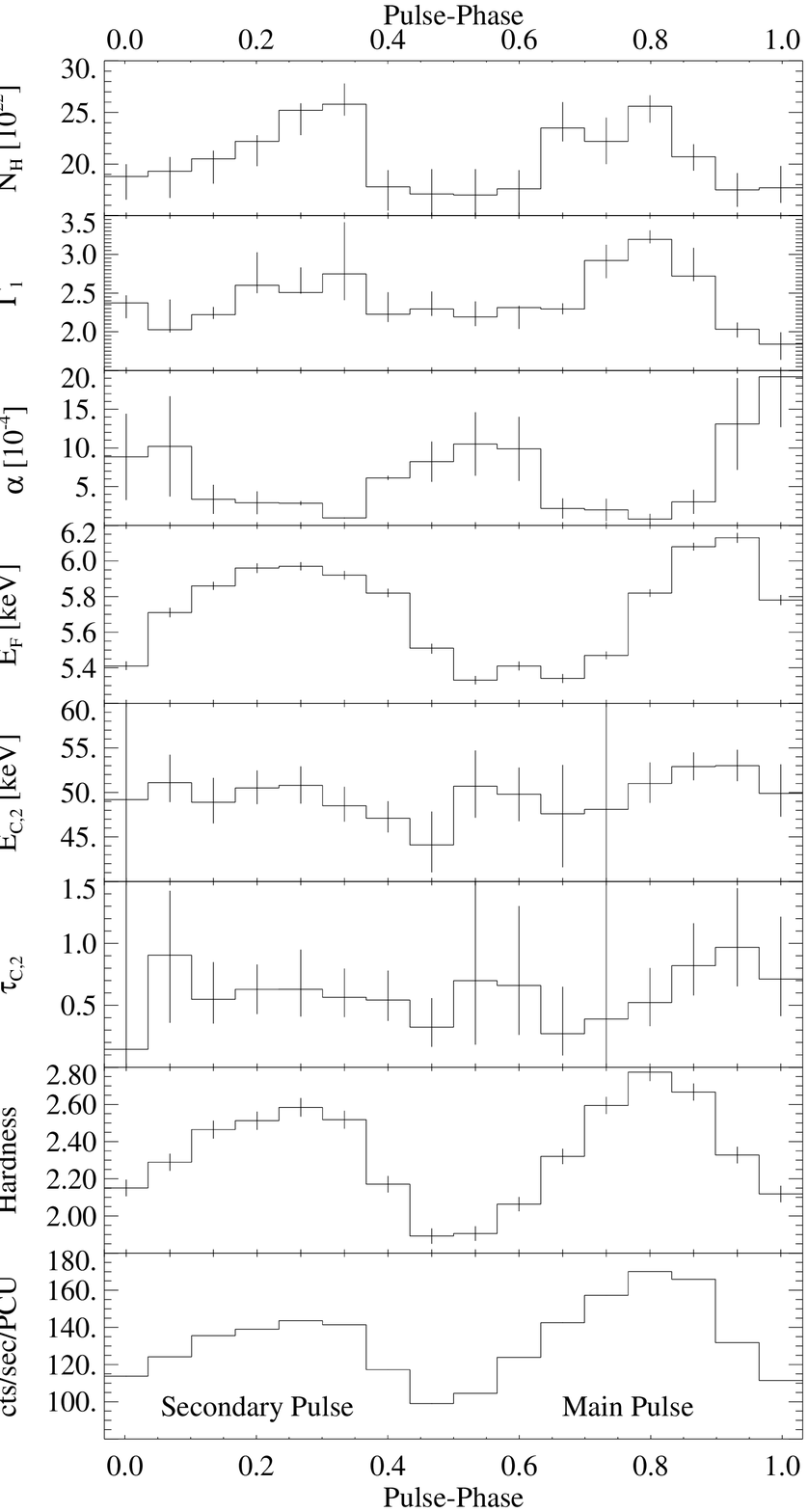}
\caption{\label{phaseparplotao3}Evolution of the fit parameters over the
  pulse for the 16~phase-bins for the 1998 data. The spectral hardness
  is defined as $R=H/S$ where $S$ is the total rate in the
  2.5--10\,\kev band and $H$ is total rate in the 15--20\,\kev band.
  The \pca-cts s$^{-1}$ PCU$^{-1}$ is in the energy range from
  15--20\,\kev and shows clearly the \vela double pulse.  $\alpha$ is
  the relative normalization of the positive power law of the
  NPEX-model and $\Gamma_1$ is the photon-index of the negative power
  law of the NPEX-model (see Eq.~\ref{npex}). $E_{\rm C,2}$ is the
  energy of the CRSF at 50\,\kev and $\tau_{\rm C,2}$ its depth. The
  width of the CRSF, $\sigma_{\rm C,2}$, has been fixed at 4\,\kev.
  See text for a discussion of the parameters. The vertical bars
  indicate the uncertainties at 90\,\% confidence level.}
\end{figure}

\begin{figure}
\includegraphics[width=\columnwidth]{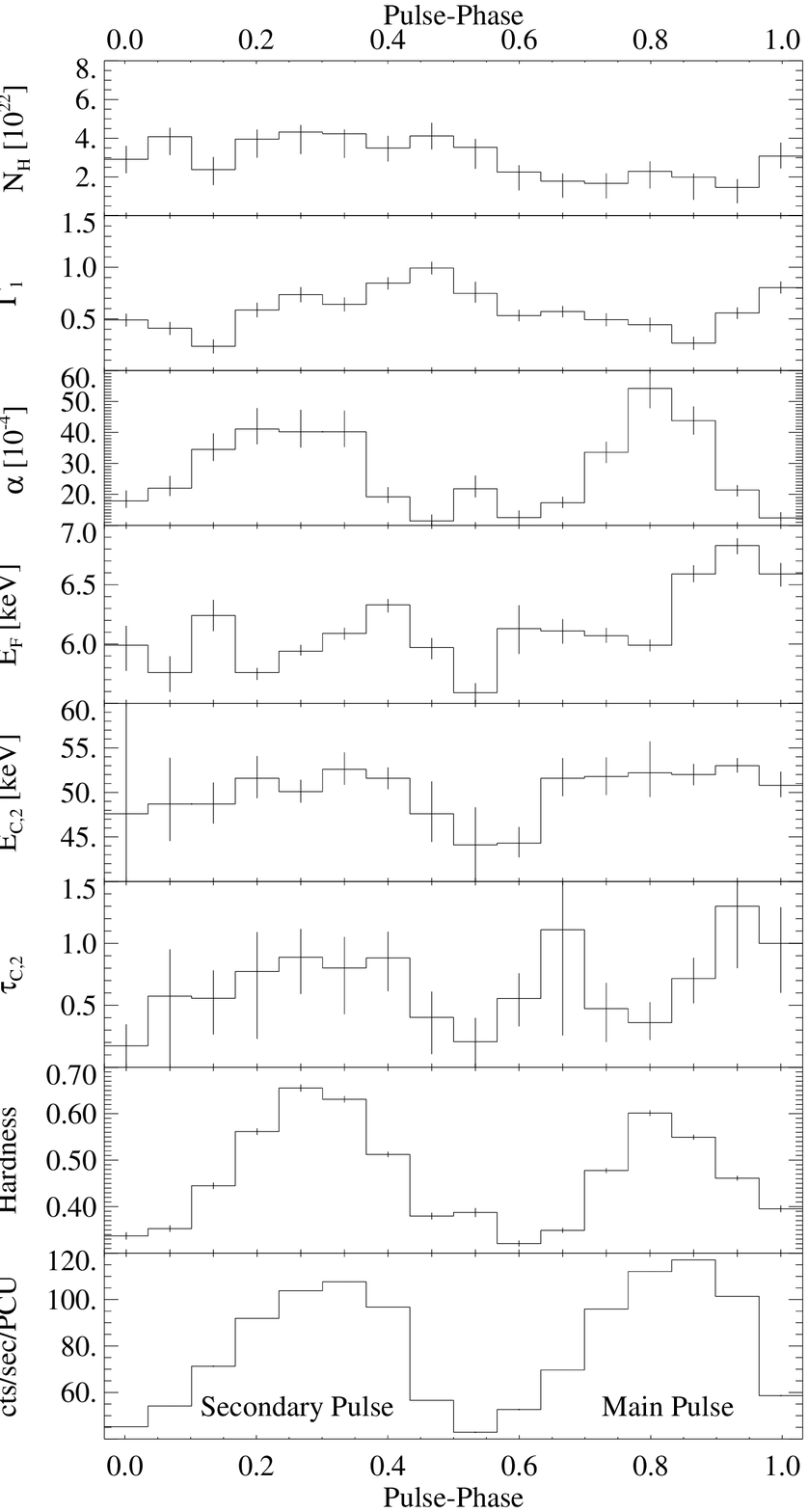}
\caption{\label{phaseparplotao4}Evolution of the fit parameters over
  the pulse for the 16~phase-bins for the 2000 data. The spectral
  hardness, the other parameters, and energy ranges are defined as in
  Fig.~\ref{phaseparplotao3}.  The vertical bars indicate the
  uncertainties at 90\,\% confidence level.}
\end{figure}

The phase resolved spectra were again modeled using the NPEX continuum
with $\Gamma_2=2$, modified by photoelectric absorption, an iron line
at 6.4\,\kev, and the CRSF at 50\,\kev.  The evolution of the relevant
spectral parameters is shown as a function of pulse phase in
Figs.~\ref{phaseparplotao3} and~\ref{phaseparplotao4} for the 1998 and
2000 data, respectively. Note that in this section, the emphasis is on
the evolution of the continuum parameters. The CRSFs will be discussed
in detail in 
the next two sections.

The spectral evolution over the pulse of the two observations is
somewhat different (see Figs.~\ref{phaseparplotao3}
and~\ref{phaseparplotao4}): in the 1998 data, \nh varies between
$15\times 10^{22}\,\text{cm}^{-2}$ and $26\times
10^{22}\,\text{cm}^{-2}$ and shows a clear correlation with flux
\citep[linear correlation coefficient $r=0.75$; ]
[Eq.~11.17]{bevington}.  In the 2000 data, \nh is lower by a factor of
10 and consistent with a constant value ($r=0.21$).

In the 1998 data, the folding energy of the NPEX-model is variable
with pulse phase: it correlates strongly with flux in the secondary
pulse (correlation coefficient $r=0.89$), while there is no such
correlation in the main pulse ($r=0.16$). In the 2000 data, this
correlation is much less pronounced: $r=0.45$ for the secondary pulse
and $r=0.14$ for the main pulse. The photon index $\Gamma_1$ of the
NPEX model varies over the pulse, but is only correlated with flux in
the 1998 data ($r=0.76$). No such correlation is present in
the 2000 data ($r=0.21$): its value in most phase bins is consistent
with 0.6, which is about a factor of 4 smaller than in the 1998 data.
Although the $\chi^2$ contours of \nh versus $\Gamma_1$ are somewhat
elongated, we note that it is impossible to fit the 2000 data with
values of $\Gamma_1$ and \nh similar to the 1998 data. Such an attempt
results in $\chi^2 > 3500$ (66 dof).  The relative normalization of
the negative power law component of the NPEX model, $\alpha$, varies
significantly over the pulse: while there is no clear correlation with
flux present in the 1998 data, the 2000 data is clearly correlated
with flux ($r=0.75$).  The hardness ratio again shows a very strong
correlation with flux in the 1998 data ($r=0.95$) and in the 2000 data
($r=0.88$).

Summarizing this section, we found that the two observations are
somewhat different. In the 1998 data, the absorption column \nh, the
photon index $\Gamma_1$, the folding energy of the NPEX model, and the
hardness ratio are correlated with flux, while all the iron line 
parameters and the relative normalization $\alpha$ of the NPEX model
are not correlated with flux. In the 2000 data, however, the relative
normalization of the NPEX model, and the hardness ratio are correlated
with flux, while the other parameters are not.

\subsubsection{The CRSFs in the 2000 data}

\begin{figure}
\includegraphics[width=\columnwidth]{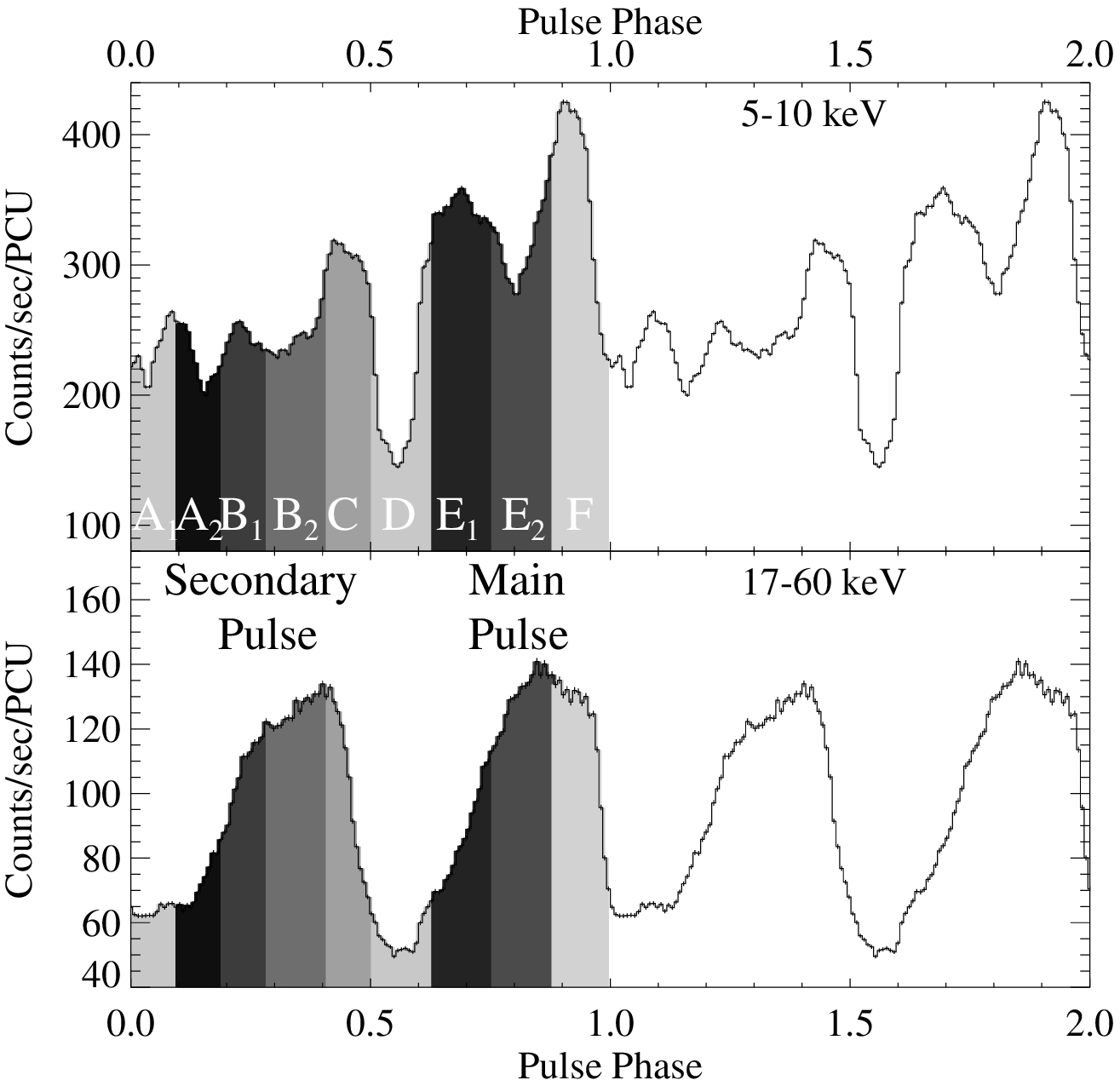}
\caption{\label{addedao4}Folded light curves (FLCs) of the 2000 data
  and definition of the nine phase bins obtained from the 2000 data.
  As in Fig.~\ref{addedao3}, we show the complex low energy pulse
  profile in the upper panel and the simpler high energy pulse profile
  in the lower panel. Due to the evolution of the pulse period, the
  FLCs of the 2000 data are somewhat offset compared to the 1998 data
  FLCs. Note that due to the higher statistics we used nine phase bins
  in the 2000 data instead of six for the 1998 data. For clarity, the folded
  light-curve is shown twice.  Note that error-bars \emph{are} shown,
  but in most cases they are too small to be seen in print.}
\end{figure}

Although the 16 phase bins provide good temporal resolution, the
relatively low exposure time per bin results in only mediocre
statistics.  Therefore we reduced the number of phase bins to obtain
better statistical quality per bin. For the 2000 data, spectra
were obtained for the rise, center, and fall of the main pulse, for
the rise, center, and fall of the secondary pulse, and for the two
pulse minima (see Fig.~\ref{addedao4} for the definition of the
individual bins).  Henceforth, we identify these phase bins with
\emph{capital} letters.

We used our standard continuum model to analyze the data; since these
sections are commited to the CRSFs, we only quote the interesting
spectral parameters here (a table of all spectral fits accompanies
this paper in electronic form only).

\begin{figure}
\includegraphics[width=\columnwidth]{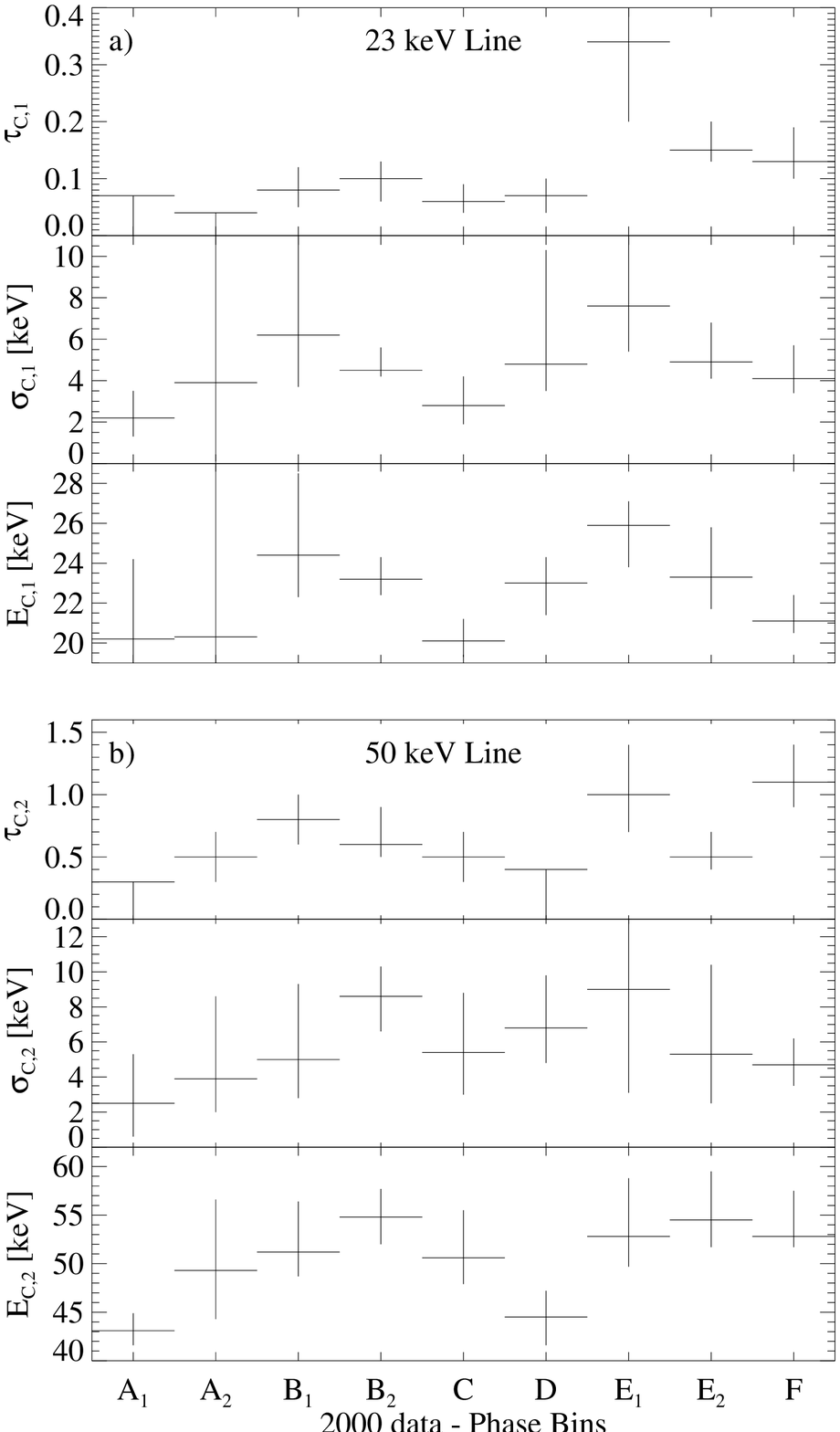}
\caption{\label{cyclphase}\textbf{a} Evolution of the depth, the
  width, and the energy of the CRSF at 23\,\kev, \textbf{b}
  evolution of the depth, the width and the energy of the CRSF at
  50\,\kev for the nine phase bins of the 2000 observation. The
  vertical bars indicate the uncertainties at the 90\,\%
  ($2.7\,\sigma$) confidence level. For discussion, see text.}
\end{figure}

The well known CRSF at 52.8\err{1.9}{1.4}\,\kev is strongest in the
fall of the main pulse with a depth of $\tau_{\rm
  C,2}=1.1\err{0.3}{0.2}$ ($F$-Test: $2.9\times 10^{-14}$).  In the
center of the main pulse (bin E$_2$), it is somewhat less deep than in
the fall ($\tau_{\rm C,2}=0.5\err{0.3}{0.1}$; see
Fig.~\ref{cyclphase}). In the rise of the main pulse (bin E$_1$) it is
again of similar depth as in the fall ($\tau_{\rm
  C,2}=1.0\err{0.4}{0.3}$, see Fig.~\ref{cyclphase}).

\begin{figure}
\includegraphics[width=\columnwidth]{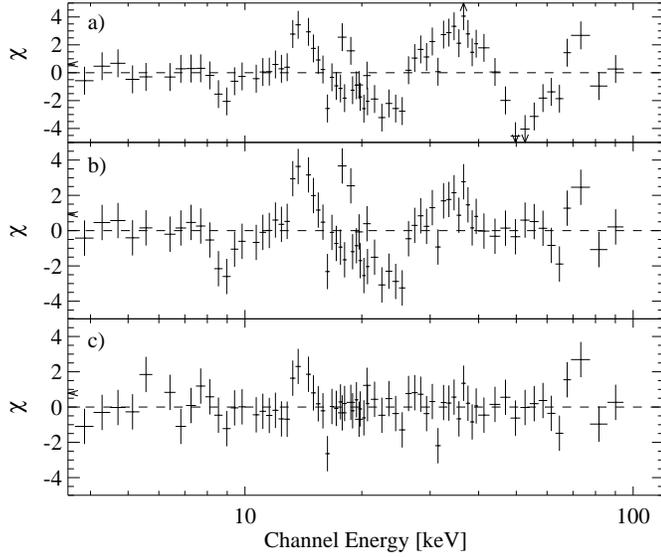}
\caption{\label{mpcao4}Residuals of fits to the spectrum of the \emph{center
    of the main pulse} (bin E$_2$) of the 2000 data. \textbf{a}
  for a model without cyclotron absorption lines, \textbf{b}
  with one line at 53.8\err{2.5}{1.9}\,\kev, and \textbf{c} with two
  lines at 23.3\err{1.3}{0.6}\,\kev and 54.5\err{5.2}{2.9}\,\kev. Note
  the feature at 23\,\kev in \textbf{b}, which we interpret as the
  fundamental cyclotron absorption line.  This line is deepest in the
  rise and most significant in the center of the main pulse and
  shallower in the other phase bins (see Fig.~\ref{spfao4} and text
  for discussion).}
\end{figure}

\begin{figure}
\includegraphics[width=\columnwidth]{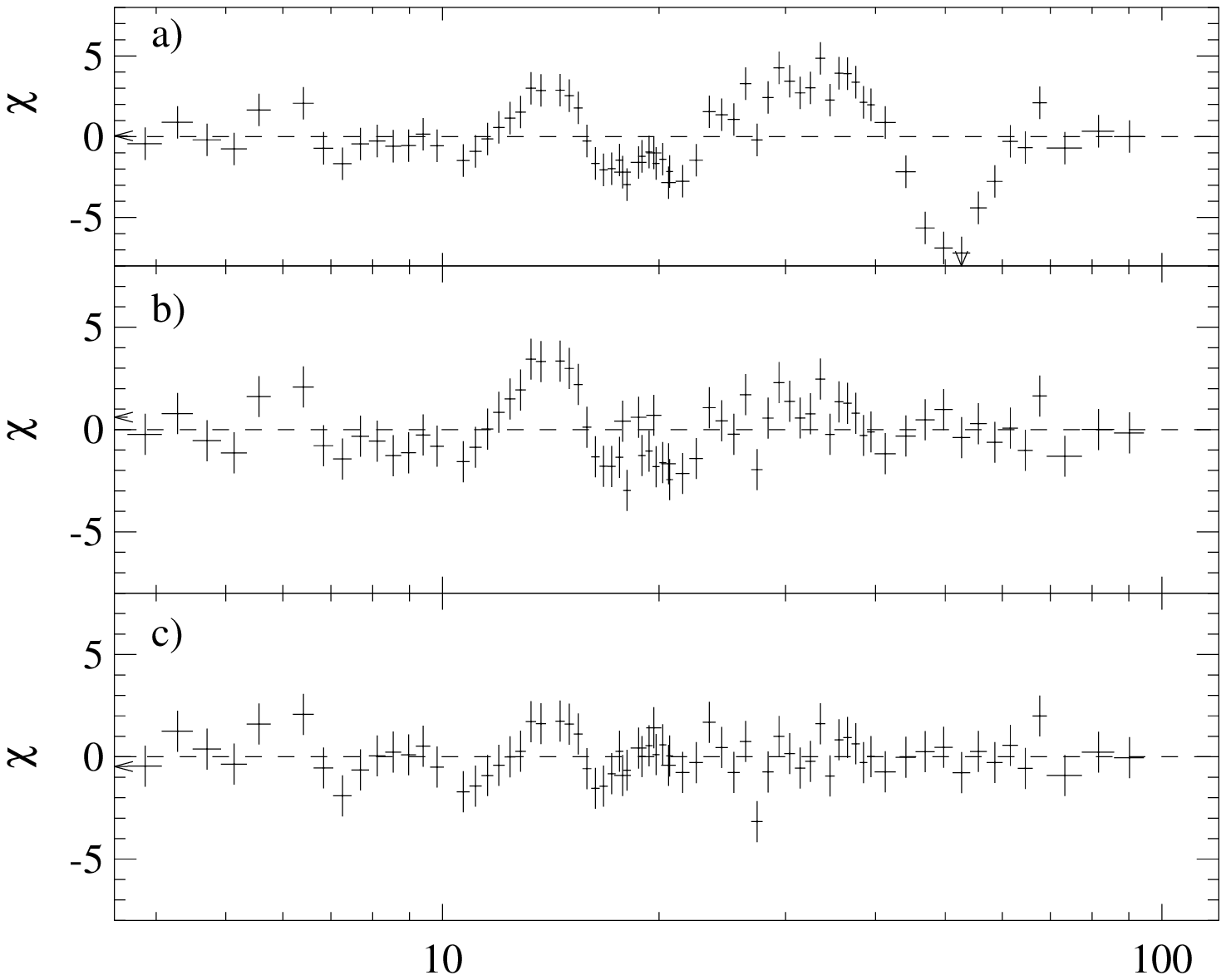}
\caption{\label{mp2ao4}\label{mpfao4}Residuals of fits to the spectrum of the fall of the main
  pulse (bin F) of the 2000 data.  \textbf{a} for a
  model without cyclotron absorption lines, \textbf{b} with one line
  at 52.8\err{1.6}{1.2}\,\kev, and \textbf{c} with two lines at
  21.1\err{0.6}{0.7}\,\kev and 52.7\err{2.0}{1.4}\,\kev.  The
  absorption feature at 23\,\kev in \textbf{a} and \textbf{b} is much
  weaker than in the center of the main pulse (see Fig.~\ref{mpcao4}),
  while the 50\,\kev line is more prominent in this phase bin than in
  any other phase bin.}
\end{figure}

\begin{figure}
\includegraphics[width=\columnwidth]{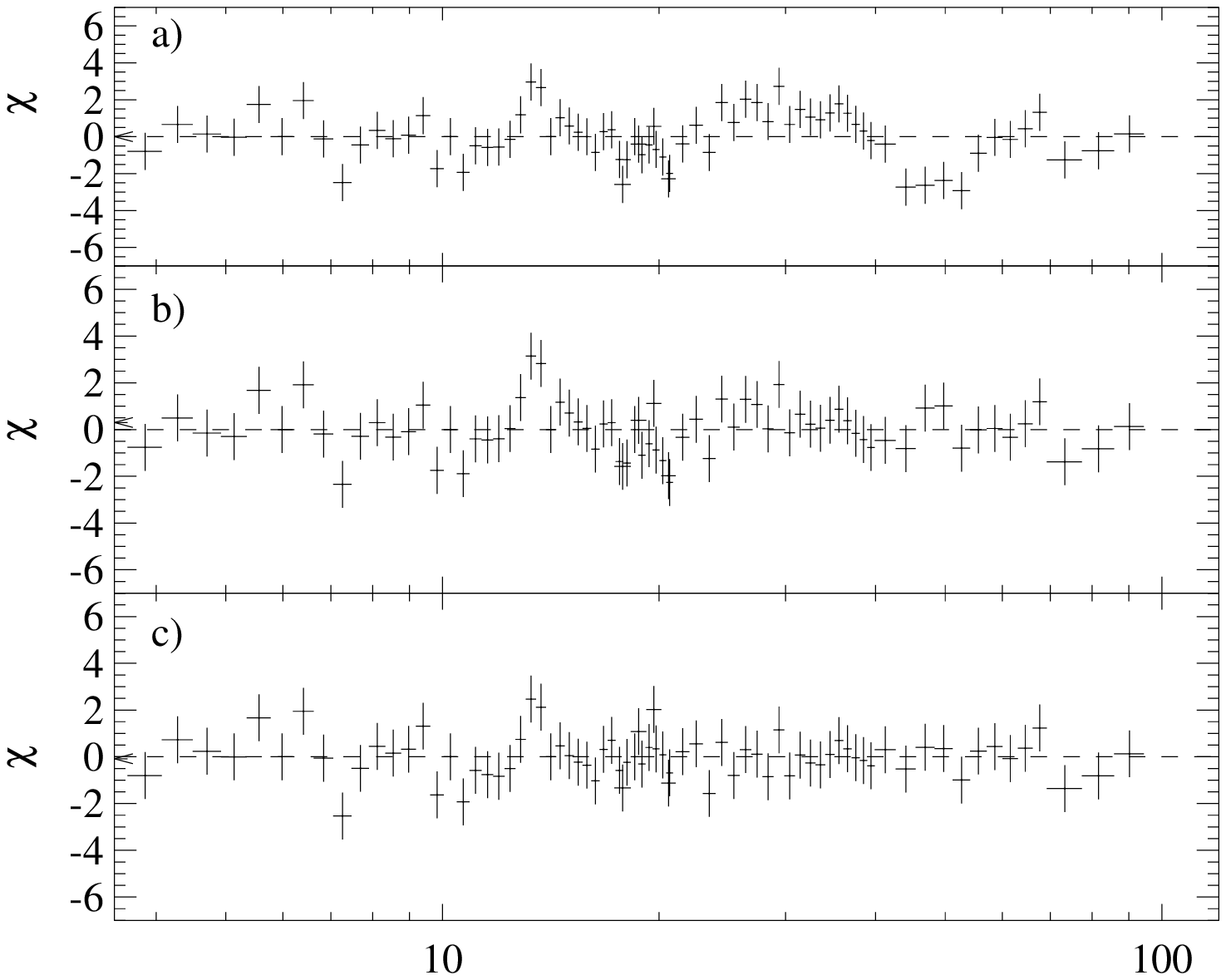}
\caption{\label{spfao4}Residuals of fits to the fall of the
  \emph{secondary} pulse (bin C) of the 2000 data. \textbf{a} 
  for a model without cyclotron absorption lines, \textbf{b} with one
  absorption line at 50.4\err{3.9}{2.5}\,\kev, and \textbf{c} with two
  lines at 20.1\err{1.4}{1.7}\,\kev and 50.5\err{6.2}{3.0}\,\kev.  The
  improvement due to the second line at 20\,\kev is very small; in
  fact, after the inclusion of the 50\,\kev line no absorption line
  feature is visible in \textbf{b}.}
\end{figure}

In the secondary pulse, the CRSF is on average less deep than in the
main pulse and the depth is -- within uncertainties -- consistent with
a constant value of 0.6 throughout the secondary pulse (bins A$_1$ to
C). In the fall (bin C) it has a depth of $\tau_{\rm
  C,2}=0.5\err{0.2}{0.2}$, but it is still very significant ($F$-Test:
$1.8\times 10^{-6}$). In the center it is of similar depth ($\tau_{\rm
  C,2}=0.6\err{0.2}{0.2}$) and also very significant ($F$-Test:
$3.7\times 10^{-11}$). In the rise (bin B$_1$) the depth $\tau_{\rm
  C,2}=0.8\err{0.3}{0.2}$ is somewhat greater, but still consistent
with a a value of 0.6.

In the pulse minima the significance of the CRSF is low (bin D,
$F$-Test: $2.8\times 10^{-2}$) or insignificant (bin A$_1$; $F$-Test:
0.59). In both cases the depth of the line is $\tau_{\rm C,2}\le0.4$
with a lower limit $<0.15$ and an upper limit of $0.5$.
Fig.~\ref{cyclphase} shows the fundamental and harmonic CRSF optical
depths in our nine phase bins of the 2000 observation.  Given these
optical depth measurements and upper limits, we find that in the main
pulse, the line is strongest on the rising and falling edges, while it
is of similar strength throughout the secondary pulse. Finally, the
line is weakest, if indeed present at all, in the pulse minima.

However, after fitting the CRSF at 50\,\kev, the fit is still not
acceptable in most phase bins: there is another absorption line like
structure present at about 25\,\kev. The most straight-forward
explanation for this feature is that it is a CRSF. After adding a
second CRSF component to our models, the resulting fits are very good.
However, as the resulting CRSFs are usually relatively shallow, we
also investigated other possibilities to fit this feature, most
notably we tried other continuum models like the Fermi-Dirac cutoff.
But these models were either completely unable to describe the data
(e.g. thermal bremsstrahlung or the cutoffpl, a power-law multiplied
by an exponential factor), known to produce artificial absorption
lines like the high energy cutoff \citep{white83a} or -- as in the
case of the Fermi-Dirac cutoff -- also produced an absorption line
like feature at 25\,\kev. The wiggle between \ca8\,\kev and 13\,\kev,
which is present in the 2000 and the 1998 data, is also present in
many other \xte observations of accreting neutron stars
\citep{coburn02a}.  It usually has the form of a small dip around
10\,\kev followed by a small hump at 12\,\kev (see for instance,
Fig.~\ref{mpfao4}); it, however, is very variable. It is present in
most phase bins (and in phase averaged spectra), but shows no clear
phase dependence. This phenomenon is yet unexplained, but it is
unrelated to the feature at 25\,\kev, since this wiggle appears in
many sources at more or less at the same energy where no CRSF is found
at 25\,\kev.  We therefore conclude that the feature at 25\,\kev is
real and interpret it as a CRSF.

The behavior of the feature at 25\,\kev is quite similar to the
50\,\kev CRSF. We found that the 25\,\kev line is most significant in
the center of the main pulse (bin E$_2$): the inclusion of a CRSF at
23.3\err{1.3}{0.6}\,\kev improves the fit significantly ($F$-Test:
$2.9\times 10^{-13}$). The depth of this line is 0.15, with a lower
limit of $0.13$. In the \emph{rise} of the main pulse (bin E$_1$), the
CRSF at 25.9\err{2.4}{1.7}\,\kev is clearly present and significant
($F$-Test: $2.8\times 10^{-7}$).  With a depth of $\tau_{\rm
  C,1}=0.33\err{0.06}{0.13}$ it is also deeper than in the center.
However, in the fall (bin F), where the 50\,\kev line is most
prominent, the fundamental line is detected with high significance
($F$-Test: $4.2\times 10^{-9}$), but it is not as deep as in the rise:
$\tau_{\rm C,1}=0.13\err{0.05}{0.02}$ (see Figs.~\ref{cyclphase}
and~\ref{mp2ao4}).  

We found that there is also a significant ($F$-Test: $ < 5\times
10^{-5}$) CRSF present at 23\,\kev throughout the secondary pulse
(B$_{1,2}$ and C). As shown in Fig.~\ref{cyclphase}, the line is of
similar depth throughout the secondary pulse: from $\tau_{\rm
  C,1}=0.06\err{0.03}{0.02}$ in the fall to $\tau_{\rm
  C,1}=0.10\err{0.03}{0.03}$ in the center, where it is also most
significant ($F$-Test: $5.6\times 10^{-10}$). In any case, the line is
somewhat shallower than in the main pulse.

In the pulse minimum \emph{after} the main pulse (bin A$_1$), an
absorption feature at 25\,\kev is not visible in the residuals and the
inclusion of a line at 22.6\err{2.9}{3.0}\,\kev does not result in an
improvement ($F$-Test: $0.29$). The upper limit for the depth of this
line is 0.10. However, in the pulse minimum \emph{before} the main
pulse (bin D), the residuals show a shallow absorption feature and the
inclusion of a CRSF at 20.2\err{1.3}{1.7}\,\kev with a depth of
$\tau_{\rm C,1}=0.07\pm0.03$ improves the fit somewhat ($F$-Test:
$3.1\times 10^{-3}$).

\begin{figure}
\includegraphics[width=\columnwidth]{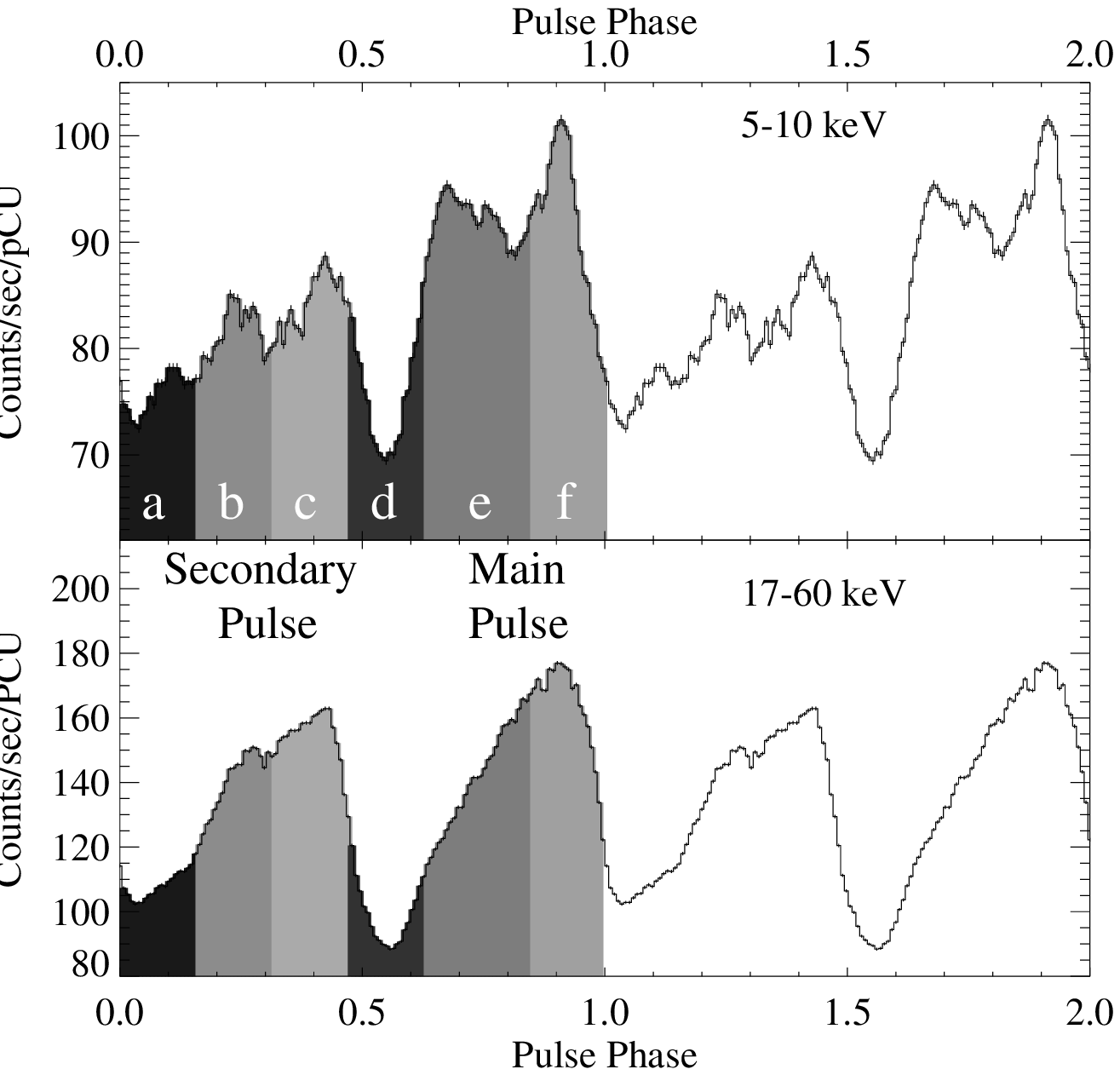}
\caption{\label{addedao3}FLCs of the 1998 data. The FLCs were generated
by converting the photon arrival times to pulse phases using the spin
period of the neutron star and then accumulating a pulse profile
binned on phase instead of time. Also shown are the definitions of the
pulse phase ranges \emph{main pulse fall}, \emph{main pulse rise},
\emph{secondary pulse fall}, \emph{secondary pulse rise}, \emph{pulse
minimum 1}, and \emph{pulse minimum 2} for the 1998 data for two
different energy bands of the \pca (for comparison with the
definitions of the 2000 data, see Fig.~\ref{addedao4}). The upper panel shows
the folded low energy pulse profile from 5 to 10\,\kev, while the
lower panel shows the high energy pulse profile above 17\,\kev.  For
clarity, the folded light-curve is shown \emph{twice}. Note that
error-bars \emph{are} shown, but in most cases they are too small to be
seen in print.}
\end{figure}

\subsubsection{The CRSFs in the 1998 data}

Since the total on-source time of the 1998 data is much less than for
the 2000 data, we used only six, broader bins in the 1998 data. We
obtained spectra for the rise and fall of the main pulse, the rise and
fall of the secondary pulse, and for the two pulse minima (see
Fig.~\ref{addedao3} for the definition of the individual bins).
Henceforth, we identify these phase bins with \emph{small} letters.

The CRSF at 52.6\err{1.4}{1.9}\,\kev is also most significant in the
fall (bin f) of the main pulse ($F$-Test: $8.2\times 10^{-11}$): its
depth is 0.8\err{0.3}{0.2}. The line is still significant in the rise
of the main pulse ($F$-Test: $1.6\times 10^{-4}$; $\tau_{\rm
  C,2}=0.4\err{0.4}{0.1}$).  Throughout the secondary pulse (bins b,
c), the line has a depth of 0.5\err{0.1}{0.1} similar to the rise of
the main pulse and is very significant ($F$-Test: $< 1.1\times
10^{-8}$). In the two pulse minima (bins a, d) the CRSF is also
present ($\tau_{\rm C,2}=0.4\err{0.1}{0.1}$), but less significant
than in the rise of the main pulse (bin e; $F$-Test: $< 2\times
10^{-3}$). Although the line depth seems to vary over the pulse, it is
within errors almost consistent with a constant value of $\tau_{\rm
  C,2}=0.6$.

\begin{figure}
\includegraphics[width=\columnwidth]{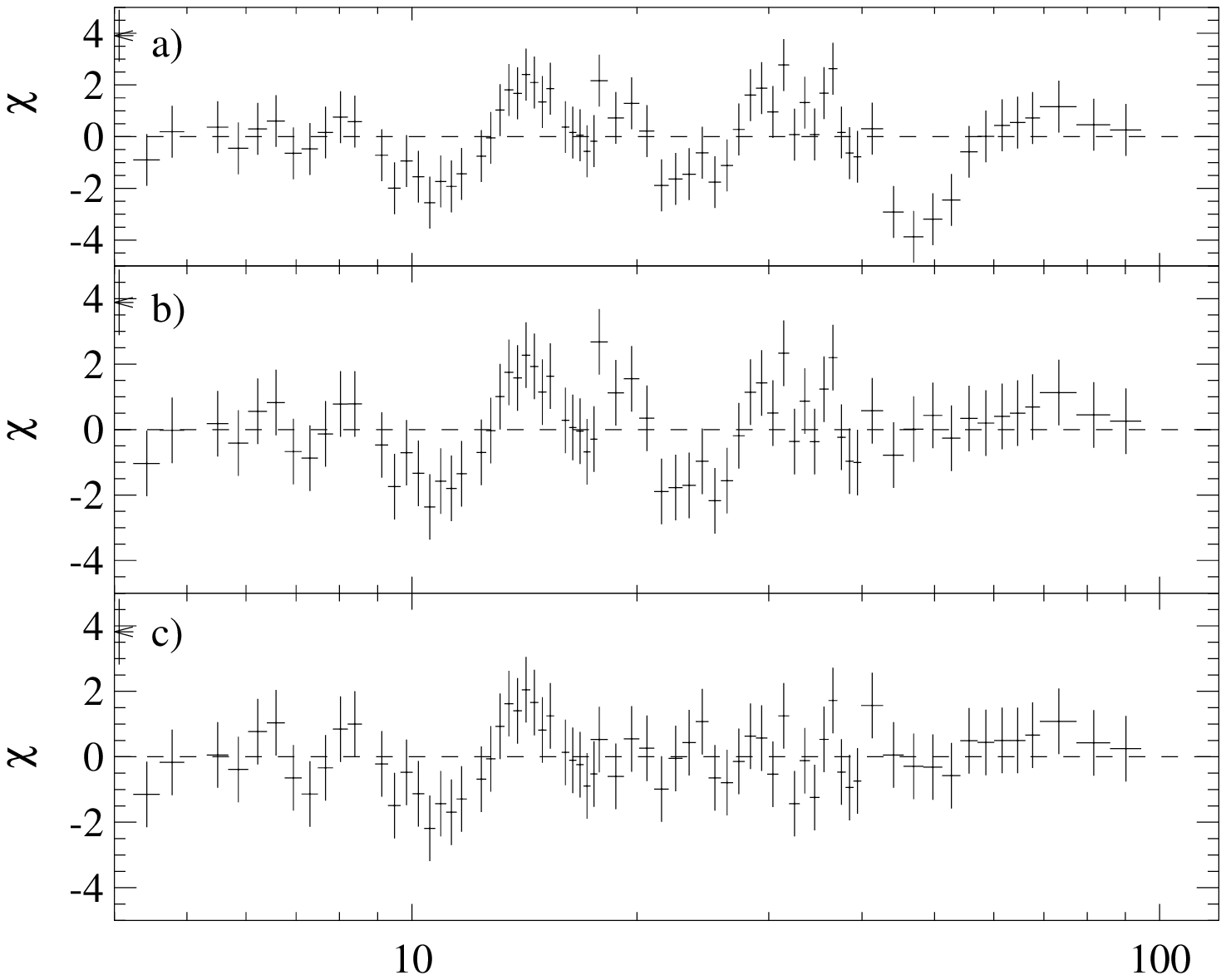}
\caption{\label{mprao3}Residuals of fits to the spectrum of the \emph{rise} of the
main pulse of the 1998 data (bin e).  \textbf{a} for a model
without cyclotron absorption lines, \textbf{b} with one line at
53.0\err{1.8}{1.4}\,\kev, and \textbf{c} with two lines at
25.1\err{1.9}{1.3}\,\kev and 53.0\err{3.8}{1.9}\,\kev. Note the
absorption feature at 25\,\kev in \textbf{b} which is not present or
much weaker in the other pulse phases.}
\end{figure}

After fitting a CRSF at 49.5\err{4.0}{3.1}\,\kev, there is still a
feature present at about 25\,\kev in the spectrum of the main pulse
\emph{rise} (bin e; see Fig.~\ref{mprao3}) -- similar to the 2000 data
(see Fig.~\ref{mpcao4}). Fitting another CRSF at
24.1\err{1.3}{0.7}\,\kev with a depth of $\tau_{\rm
  C,1}=0.08\err{0.03}{0.02}$ improves the fit significantly ($F$-test
probability for the CRSF being a chance improvement: $6.3\times
10^{-5}$).

In the fall of the main pulse there is also an indication of the
presence of an absorption feature at 25\,\kev after fitting the
52.7\err{1.6}{1.1}\,\kev CRSF.  Fitting another cyclotron absorption
line at 23.6\err{1.0}{2.2}\,\kev with a depth of $\tau_{\rm
C,1}=0.06\err{0.01}{0.02}$ improves the fit with an $F$-test
probability of $2.7\times 10^{-3}$.

However, between the two pulses we found no significant line at
24\,\kev: in the pulse minimum 1 (bin a) and pulse minimum 2 (bin d),
the inclusion of a CRSF at 23.9\err{2.1}{1.9}\,\kev resulted in no
significant improvement; upper limits for the depth of the line are
0.06 (pulse minimum 1) and 0.05 (pulse minimum 2).  The same applies
for the spectra of the rise and fall of the secondary pulse (bins b
and c): the inclusion of a CRSF at 24\,\kev does \emph{not} improve
the fit significantly ($F$-Test: $> 9.8\times 10^{-2}$).  Note that
this does not exclude the presence of a CRSF at 24\,\kev, it is just
not detected: the upper limit for the depth of the CRSF at
24\,\kev is $0.04$ for these phase bins.

\begin{table*}
\caption{\label{addedtab}Fit results for some selected phase bins from
  the 2000 data. All uncertainties quoted in this paper are at the 90\,\%
  ($2.7\,\sigma$) confidence 
  level. Note that there is an improvement when using a line at
  25\,\kev also in the secondary pulse but this improvement is very
  small compared to the improvement in the center of the main
  pulse (see below).  The complete tables for the 1998 and 2000 data are
  available in  electronic form only.}

\begin{tabular}{lrrrrrrrrr}
\hline
& \multicolumn{3}{c}{Main Pulse Fall (F)} & \multicolumn{3}{c}{Secondary
  Pulse Fall (C)} & \multicolumn{3}{c}{Main Pulse Center (E$_2$)} \\
& w/o cycl. & 1 cycl. &  2 cycl. & w/o cycl. & 1 cycl. & 2 cycl. & w/o
cycl. & 1 cycl. & 2 cycl.\\ 
\hline
\nh $[10^{22}] $& 2.4\err{0.5}{0.8} & 2.5\err{0.4}{1.8} &
2.4\err{0.9}{0.6} & 4.2\err{0.5}{0.9} & 4.2\err{0.2}{1.3} &
4.2\err{0.6}{0.8} & 1.9\err{0.5}{0.9} & 1.8\err{0.3}{1.3} &
2.9\err{0.4}{1.2} \\
$\Gamma_1$ & 0.72\err{0.04}{0.07} & 0.61\err{0.06}{0.06} &
0.81\err{0.12}{0.08} & 1.04\err{0.04}{0.08} & 0.98\err{0.06}{0.07} &
1.07\err{0.06}{0.08} & 0.36\err{0.06}{0.07} & 0.29\err{0.07}{0.07} &
0.58\err{0.10}{0.11} \\
$\Gamma_2$ & $2$ & $2$ & $2$ & $2$ & $2$ & $2$ & $2$ & $2$ & $2$ \\
\efold & 6.36\err{0.03}{0.03} & 6.38\err{0.09}{0.06} &
6.37\err{0.08}{0.07} & 5.95\err{0.04}{0.03} & 6.14\err{0.09}{0.06} &
6.04\err{0.08}{0.07} & 6.14\err{0.03}{0.03} & 6.28\err{0.04}{0.04} &
6.02\err{0.17}{0.05} \\
$E_{\rm C,1}$ [\kevnxs] & -- & -- & 21.1\err{0.7}{0.7} & -- & -- &
20.1\err{1.3}{1.6} & -- & -- & 23.3\err{1.3}{0.6} \\
$\sigma_{\rm C,1}$ [\kevnxs] & -- & -- & 4.1\err{1.6}{0.7} & -- &
-- & 2.8\err{1.4}{0.9} & -- & -- & 4.9\err{1.9}{0.8} \\
$\tau_{\rm C,1}$ & -- & -- & 0.13\err{0.05}{0.02} & -- & -- &
0.06\err{0.03}{0.02} & -- & -- & 0.15\err{0.07}{0.02} \\
$E_{\rm C,2}$ [\kevnxs] & -- & 53.0\err{1.6}{1.2} &
52.8\err{1.9}{1.4} & -- & 50.5\err{3.9}{2.4} & 50.6\err{4.9}{2.7} &
-- & 53.8\err{2.3}{1.9} & 54.5\err{5.0}{2.8} \\
$\sigma_{\rm C,2}$ [\kevnxs] & -- & 4.9\err{1.2}{0.9} &
4.7\err{1.5}{1.2} & -- & 4.9\err{2.6}{2.5} & 5.4\err{3.4}{2.4} & --
& 3.7\err{1.6}{1.6} & 5.3\err{5.1}{2.8} \\
$\tau_{\rm C,2}$ & -- & 1.3\err{0.3}{0.2} & 1.1\err{0.3}{0.2} & -- &
0.6\err{0.4}{0.2} & 0.5\err{0.2}{0.2} & -- & 0.8\err{0.5}{0.2} &
0.5\err{0.3}{0.1} \\
\hline
$\chi^2 (\text{dof})$ & 424 (64) & 144 (61) & 62 (58) & 121 (64) & 75
(61) & 50 (58) & 258 (64) & 174 (61) & 61 (58) \\
\hline
\end{tabular}
\end{table*} 

\section{Summary and Discussion}
\label{discussion}
\subsection{The Pulse Profile}
\label{discussion_profile}
Despite the strong pulse to pulse variations and the overall strong
variability of \vela including flaring activity (see
Fig.~\ref{ao3light}, \ref{ao4light}) and ``off-states''
\citep{kreykenbohm99a,inoue84a}, the pulse profile at higher
\emph{and} lower energies is remarkably stable \citep[compare the
pulse profile in Fig.~\ref{profileao4} and e.g.][]{mcclintock76a}.
Therefore, the low energy profile is not due to any random short term
fluctuations. 

Such an evolution of the pulse profile and therefore the spectral
shape over the X-ray pulse is typical for accreting X-ray pulsars,
however, its interpretation is difficult.  While the double pulse at
energies above 20\,\kev can be attributed to the two magnetic poles
\citep{raubenheimer90a}, there is no such straightforward explanation
for the low energy pulse profile.  

Early explanations of the pulse profile invoked 
varying photoelectric absorption over the pulse
\citep{nagase83a}. In this picture, for example, the ``dip'' at
pulse phase $\sim$0.8 is attributed to the accretion column passing
through the line of sight.  However, there is no such straightforward
explanation for the complex profile of the secondary pulse and
furthermore, our phase resolved spectra do \emph{not} show an
increased absorption in the center of the main pulse
(Fig.~\ref{phaseparplotao4}). It therefore seems more plausible to
attribute the variation of the spectral parameters with pulse phase to
the mechanism generating the observed radiation, such as anisotropic
propagation of X-rays in the magnetized plasma of the accretion column
\citep{nagel81b} or the influence of the magnetic field configuration
\citep{mytrophanov78a}. Due to the uncertainties associated with
modeling the accretion column (see Sect.~\ref{models}), only the
general shape of the pulse profile can be described with these models,
and the substructures cannot be explained.  A good description of the
processes responsible for the complex shape of the pulse profile is
thus still missing.

We note, however, that in the strong magnetic field of the accretion
column the motion of electrons perpendicular to the magnetic field is
constrained, while they move freely parallel to the field lines.
As a result, one expects the thermal motion of the
electrons to produce a harder spectrum when viewed parallel to the
$B$-field than when viewed perpendicular to the
field. In a pencil beam geometry, one would then expect the pulse
maxima to have a harder spectrum than the minima, similar to what has
been seen in Vela~X-1.

\subsection{The existence of the 25\,\kev line}\label{sec:existence}
It is very unlikely that the 25\,\kev feature results from a
calibration problem. Fits to Crab spectra taken close to the 2000
observation have shown that there are no deviations in the relevant
area between 20\,\kev and 30\,\kev in the \hexte (see
Fig.~\ref{crabsys}).  Furthermore, we have limited the energy range of
the \pca to $\le$21\,\kev, so the detection of the line is almost only
due to the \hexte. Finally, we note that the line is only visible in
one pulse phase of the 1998 data, which should be subject to similar
calibration uncertainties as the 2000 data. Since the \hexte has a
continuous automatic gain control, the \hexte response is very stable
throughout the mission, and thus the calibration is the same for AO1 
through AO4.

Our observations indicate that the feature is quite variable with
pulse phase. As discussed in the previous sections, the depth of the
line varies strongly with pulse phase. In certain phase bins, the
lower limit of the depth is 0.20, while in another phase bin its upper
limit is 0.09. This variability also makes it highly unlikely that the
25\,\kev feature is due to a calibration problem as it would then be
present in all phase bins.  Apart from that, the feature appears
at about half the energy of the 50\,\kev cyclotron line: the lower
line is at 22.9\err{2.9}{0.9}\,\kev, while the second line is at
50.9\err{0.6}{0.7}\,\kev (2000 data, phase-averaged spectrum). This
results in an average coupling factor of $2.15\pm0.19$, in agreement
with a factor of 2.0 expected for a first harmonic. We also note that
in the main pulse rise (2000 data) where the 25\,\kev CRSF is deepest,
we find a coupling factor of almost \emph{exactly} 2.0. However, we
caution that although theoretical models predict a value of 2.0
\citep[][and references therein]{harding91b}, recent observations have
shown that the fundamental line can be heavily distorted \citep[as in
4U\,0115+63;][]{heindl99b,santangelo99b} and therefore a coupling
factor of exactly 2.0 is not necessarily to be expected.

Furthermore, the 25\,\kev line varies with time: It was barely visible
in the 1998 data and was not seen at all by \sax \citep{orlandini97c}.
Adding to that, the strong phase variability of the line
makes it difficult to detect.  Therefore it is not surprising that
\citet{orlandini97c} and \citet{kreykenbohm00b} did not detect this
line in their phase-averaged spectra. Using phase resolved \ginga
data, \citet{mihara95a} also reported a feature at 24\,\kev in the
main pulse while it was insignificant outside the main pulse.

\begin{figure}
\includegraphics[width=\columnwidth]{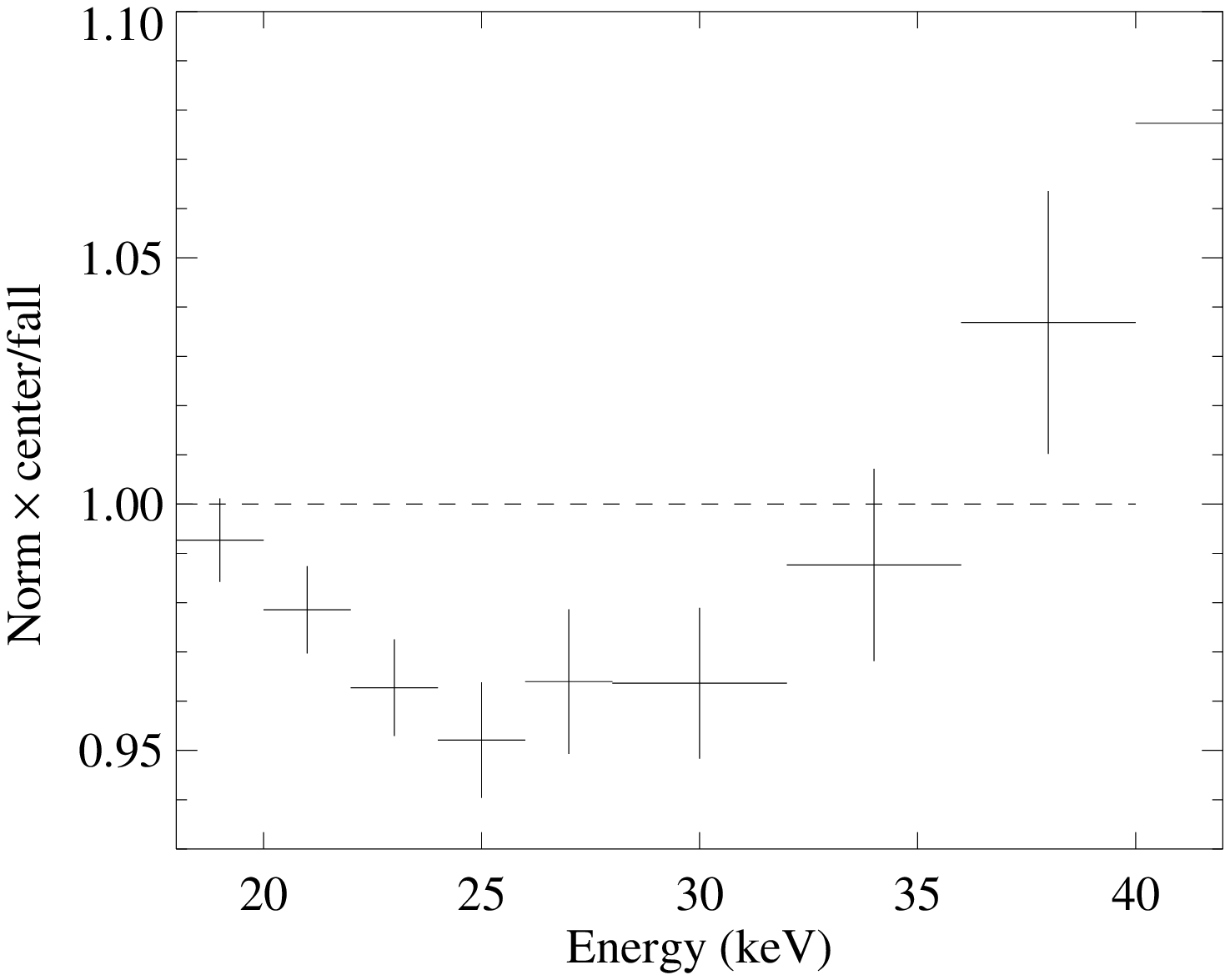}
\caption{\label{divide}Raw \hexte count rate spectrum of the
  center of the main pulse divided by the raw \hexte count rate spectrum of
  the fall of the secondary pulse using the 2000 data. Note the clear
  lack of counts between 20\,\kev and 30\,\kev in the spectrum of the
  center of the main pulse. }
\end{figure}

Finally, to be completely independent of any applied model or response
matrices, we divided the raw count rate spectrum of the center of the
main pulse by the count rate spectrum of the fall of the secondary
pulse. The result is shown in Fig.~\ref{divide}: there is a clear dearth
of counts between 20\,\kev and 30\,\kev, just where we find the
25\,\kev line with about the same width. This means that there is an
absorption line like feature in the raw count rate spectrum of the
center of the main pulse -- independent of the applied model or the
response matrix.

In summary, we have shown that there is an absorption line like
feature at 25\,\kev in some phase bins, while it is shallower or far
less significant in other phase bins. The most straightforward
interpretation of this feature is that it is the fundamental CRSF,
while the 50\,\kev line is the first harmonic.

\subsection{CRSFs in accreting pulsars}

\begin{figure}
\includegraphics[width=\columnwidth]{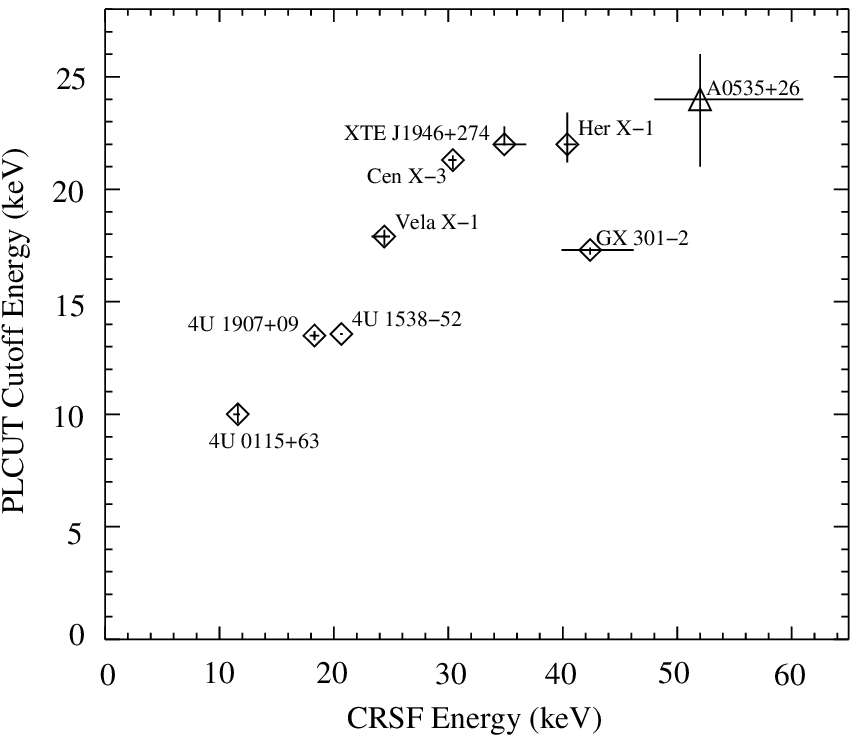}
\caption{\label{wayne}Correlation between the centroid energy of the
  fundamental CRSF and the cutoff energy (when using a power law/high
  energy cutoff model instead the NPEX model), based on \xte data
  (with the exception of A\,0535$+$26). See \citet{coburn01b} for a
  complete discussion.  With a line energy of 25\,\kev, \vela fits
  nicely into the sequence, whereas it would not fit in if the energy
  of the fundamental line were 50\,\kev. }
\end{figure}

There are now about a dozen X-ray pulsars for which cyclotron lines
have been detected. Most of these sources have been observed with
\xte, such that a systematic study of the properties of these sources
and their CRSFs starts to become feasible. Although such a study is
outside the scope of this paper, we want to point out that our
detection of a CRSF at $\sim$25\,keV in \vela is consistent with the
general properties of CRSF sources. In their study of CRSFs from
\ginga and other instruments, \citet{makishima99a} found a correlation
between the energy of the fundamental CRSF and the cutoff energy in
the canonical \citet{white83a} X-ray pulsar continuum.  Recently, this
correlation has been reinvestigated by \citet{coburn01b} using \xte
data from all sources with known CRSFs. Since \xte has a broader
energy range than previous missions, such a study is able to cover a
wider range of CRSF energies.  \citet{coburn01b} finds that most \xte
observed pulsars follow the \citet{makishima99a} correlation up to
35\,\kev where a roll-over is observed \citep[see Fig.~\ref{wayne} and
][ for a discussion]{coburn01b}.

The only exceptions so far are GX 301$-$2 which shows very high
amplitude \nh variations which make the determination of the continuum
uncertain, 4U\,1626$-$67 follows the correlation only in certain pulse
phases, while X Per has a very different continuum \citep{coburn01a} and
therefore a comparison with the other sources is difficult if possible
at all.

We note that \vela fits nicely the correlation of Fig.~\ref{wayne},
provided the fundamental line is at $\sim$25\,keV and not at
50\,\kev. 

As we have pointed out in section~\ref{sec:existence}, the fundamental
line in \vela is weak and only detectable in some specific pulse
phases. This is similar to the other three sources where more than one
fundamental CRSF has been claimed so far: 4U\,0115$+$63
\citep{heindl99b,santangelo99b}, A\,0535$+$26
\citep{kendziorra94a,maisack97a}, and 4U\,1907$+$07
\citep{cusumano98a}.  In 4U\,0115$+$63, the fundamental CRSF is quite
complex and decidedly non-Gaussian in shape -- \citet{heindl99b}
require two overlapping lines in order to model this line, while the
higher 4 harmonics are much more Gaussian in shape. Here, all lines
show strong variation with pulse phase, but are almost always present.
For A\,0535$+$26, a very prominent line at \ca100\,\kev has been
reported by \citet{maisack97a} using \osse, while a 50\,\kev feature
presented by \citet{kendziorra94a} from \hexe data was quite weak and
difficult to model. Finally, \citet{cusumano98a} find in
4U\,1907$+$09, that the line at 39\,\kev is again very prominent while
its $\sim$19\,\kev fundamental is weak\footnote{Recently, a CRSF
  feature at $\sim$70\,\kev has also been claimed for Her~X-1 (Segreto
  et al., 2002, in prep.). If this proves to be true, Her~X-1 would be
  the only object with more than one cyclotron line in which the
  fundamental line is stronger than the upper harmonics.}.

Such a complex line shape of the fundamental line is consistent with
Monte Carlo simulations of the generation of cyclotron lines in
accreting X-ray pulsars \citep{isenberg97a,araya99a,araya00a}. These
simulations show that the shape of the CRSF depends strongly on the
assumed geometry, electron temperature, and optical depth of the
plasma. It has been pointed out by \citet{araya99a}, that under
certain viewing angles, the fundamental line is not a simple
absorption line, but has a very complex shape that includes, for
example, ``P~Cygni'' like emission wings. These emission wings are
caused by the angular dependence of the scattering cross section,
multiple resonant scattering, and by higher order effects such as
``photon spawning'' \citep{araya00a}. 

The feature seen around 20\,\kev (Figs.~\ref{mpcao4}, \ref{mpfao4}, and
\ref{spfao4}) might be such an emission wing. Fitting this feature
(where present) with a Gaussian produces a fit comparable to fitting
an absorption line. The resulting Gaussians are relatively narrow:
0.7\err{0.6}{0.7}\,\kev in the fall of the secondary pulse and
1.5\err{0.2}{0.4}\,\kev in the fall of the main pulse. The Gaussian is
broadest in the rise of the main pulse with a width of
3.0\err{1.0}{0.7}\,\kev. Due to this narrowness we believe that this
feature is a wing of a CRSF and not due to improper modeling of the
continuum, e.g. due to a Wien bump between 10\,\kev and 20\,\kev which
would have to be much broader.

Higher order CRSFs tend to be much less affected by these phenomena.
When such complicated line profiles are convolved with the moderate
energy resolution of todays X-ray detectors, it is possible that such
emission wings result in line profiles where the line is almost
undetectable or where the line shows the shallow non-Gaussian shape
seen in 4U~0115$+$63.  Since the strength of the wings depends on the
inclination angle, one could envisage a scenario where the geometry of
the accretion column is such that the fundamental is completely filled
in except for certain pulse phases, explaining the difficulty of
detecting the fundamental CRSF in these systems.  Admittedly, these
simulations have only begun and the computation of larger grids of
simulated and more realistic CRSFs for observational work has only
recently become computationally feasible.  However, if this
speculation were true, it would provide a natural explanation for the
scarcity of X-ray pulsars where more than one CRSF is observable.

\begin{acknowledgements}
  This work has been financed by a grant from the DLR, NASA grant
  NAG5-30720, NSF travel grant INT9815741, and a travel grant from
  the DAAD.
\end{acknowledgements}

\bibliographystyle{aa}
\bibliography{mnemonic,aa_abbrv,velax1,div_xpuls,xpuls,cyclotron,books,roentgen,satelliten,foreign}

\begin{thebibliography}{61}
\expandafter\ifx\csname natexlab\endcsname\relax\def\natexlab#1{#1}\fi

\bibitem[{Alexander {et~al.}(1996)Alexander, Davila, \&
  Dimattio}]{alexander96a}
Alexander, S.~G., Davila, J., \& Dimattio, D.~J. 1996, ApJ, 459, 666

\bibitem[{Araya \& Harding(1999)}]{araya99a}
Araya, R.~A. \& Harding, A.~K. 1999, ApJ, 517, 334

\bibitem[{Araya-G{\'o}chez \& Harding(2000)}]{araya00a}
Araya-G{\'o}chez, R.~A. \& Harding, A.~K. 2000, ApJ, 544, 1067

\bibitem[{Arnaud(1996)}]{arnaud96a}
Arnaud, K.~A. 1996, in Astronomical Data Analysis Software and Systems {V}, ed.
  J.~H. Jacoby \& J.~Barnes, ASP Conf. Ser. 101, San Francisco, 17

\bibitem[{Bevington \& Robinson(1992)}]{bevington}
Bevington, P.~R. \& Robinson, D.~K. 1992, Data Reduction and Error Analysis for
  the Physical Sciences (Boston: McGraw-Hill)

\bibitem[{Bradt {et~al.}(1993)Bradt, Rothschild, \& Swank}]{bradt93a}
Bradt, H.~V., Rothschild, R.~E., \& Swank, J.~H. 1993, A\&AS, 37, 355

\bibitem[{Brainerd \& {M\'esz\'aros}(1991)}]{brainerd:91a}
Brainerd, J.~J. \& {M\'esz\'aros}, P. 1991, ApJ, 369, 179

\bibitem[{Burnard {et~al.}(1991)Burnard, Arons, \& Klein}]{burnard91a}
Burnard, D.~J., Arons, J., \& Klein, R.~I. 1991, ApJ, 367, 575

\bibitem[{Choi {et~al.}(1996)Choi, Dotani, Day, \& Nagase}]{choi96a}
Choi, C.~S., Dotani, T., Day, C. S.~R., \& Nagase, F. 1996, ApJ, 471, 447

\bibitem[{Coburn(2001)}]{coburn01b}
Coburn, W. 2001, {Ph.D. T}hesis, University of California, San Diego

\bibitem[{Coburn {et~al.}(2001)Coburn, Heindl, Gruber, Rothschild, Staubert,
  Wilms, \& Kreykenbohm}]{coburn01a}
Coburn, W., Heindl, W.~A., Gruber, D.~E., {et~al.} 2001, ApJ, 552, 738

\bibitem[{Coburn {et~al.}(2002)Coburn, Heindl, Rothschild, Gruber, Kreykenbohm,
  Wilms, Kretschmar, \& Staubert}]{coburn02a}
Coburn, W., Heindl, W.~A., Rothschild, R.~E., {et~al.} 2002, ApJ, submitted

\bibitem[{Cusumano {et~al.}(1998)Cusumano, {Di Salvo}, Burderi, Orlandini,
  Piraino, Robba, \& Santangelo}]{cusumano98a}
Cusumano, G., {Di Salvo}, T., Burderi, L., {et~al.} 1998, A\&A, 338, L79

\bibitem[{Davidson \& Ostriker(1973)}]{davidson73a}
Davidson, K. \& Ostriker, J.~P. 1973, ApJ, 179, 585

\bibitem[{Gruber {et~al.}(1996)Gruber, Blanco, Heindl, Pelling, Rothschild, \&
  Hink}]{gruber96a}
Gruber, D.~E., Blanco, P.~R., Heindl, W.~A., {et~al.} 1996, A\&AS, 120, 641

\bibitem[{Haberl \& White(1990)}]{haberl90a}
Haberl, F. \& White, N.~E. 1990, ApJ, 361, 225

\bibitem[{Harding(1994)}]{harding94a}
Harding, A.~K. 1994, in Proc. conf. The Evolution of {X}-ray Binaries, ed.
  S.~Holt \& C.~S. Day, AIP Conference Proceedings 308, 429

\bibitem[{Harding \& Daugherty(1991)}]{harding91b}
Harding, A.~K. \& Daugherty, J.~K. 1991, ApJ, 374, 687

\bibitem[{Heindl {et~al.}(1999)Heindl, Coburn, Gruber, Pelling, Rothschild,
  Wilms, Pottschmidt, \& Staubert}]{heindl99b}
Heindl, W.~A., Coburn, W., Gruber, D.~E., {et~al.} 1999, ApJ, 521, L49

\bibitem[{Hua \& Titarchuk(1995)}]{hua:95a}
Hua, X.-M. \& Titarchuk, L. 1995, ApJ, 449, 188

\bibitem[{Inoue {et~al.}(1984)Inoue, Ogawara, Ohashi, \& Waki}]{inoue84a}
Inoue, H., Ogawara, Y., Ohashi, T., \& Waki, I. 1984, PASJ, 36, 709

\bibitem[{Isenberg {et~al.}(1998)Isenberg, Lamb, \& Wang}]{isenberg97a}
Isenberg, M., Lamb, D.~Q., \& Wang, J. C.~L. 1998, ApJ, 505, 688

\bibitem[{Jahoda {et~al.}(1996)Jahoda, Swank, Giles, Stark, Strohmayer, Zhang,
  \& Morgan}]{jahoda96a}
Jahoda, K., Swank, J.~H., Giles, A.~B., {et~al.} 1996, in {EUV}, {X-ray}, and
  Gamma-Ray Instrumentation for Astronomy {VII}, Proc. {SPIE}, ed. O.~H.
  Siegmund \& M.~A. Gummin, Vol. 2808, SPIE, 59--70

\bibitem[{Kendziorra {et~al.}(1994)Kendziorra, Kretschmar, Pan, Kunz, Staubert,
  Pietsch, Tr{\"u}mper, Efremov, \& Sunyaev}]{kendziorra94a}
Kendziorra, E., Kretschmar, P., Pan, H.~C., {et~al.} 1994, A\&A, 291, L31

\bibitem[{Kendziorra {et~al.}(1992)Kendziorra, Mony, Kretschmar, Maisack,
  Staubert, D\"obereiner, Englhauser, Pietsch, Reppin, Tr\"umper, Efremov,
  Kaniovski, \& Sunyaev}]{kendziorra92a}
Kendziorra, E., Mony, B., Kretschmar, P., {et~al.} 1992, in Frontiers of
  {X}-Ray Astronomy, Proc. {XXVIII} Yamada Conf., ed. Y.~Tanaka \& K.~Koyama,
  Frontiers Science Series, 2, 51--52

\bibitem[{Kretschmar {et~al.}(1999)Kretschmar, Kreykenbohm, Wilms, Staubert,
  Heindl, Gruber, \& Rothschild}]{kretschmar99a}
Kretschmar, P., Kreykenbohm, I., Wilms, J., {et~al.} 1999, in Proc. 5th Compton
  Symposium, ed. M.~McConnell \& J.~Ryan, AIP Conf. Proc. No. 510, 163

\bibitem[{Kretschmar {et~al.}(1997)Kretschmar, Kreykenbohm, Wilms, Staubert,
  Maisack, Kendziorra, Heindl, Rothschild, Gruber, \& Grove}]{kretschmar96b}
Kretschmar, P., Kreykenbohm, I., Wilms, J., {et~al.} 1997, in The Transparent
  Universe, Proc. 2nd {INTEGRAL} Workshop, ed. C.~Winkler, T.~J.-L.
  Courvoisier, \& P.~Durouchoux, ESA SP 382, Noordwijk, 141--144

\bibitem[{Kretschmar {et~al.}(1996)Kretschmar, Pan, Kendziorra, Maisack,
  Staubert, Pietsch, {Tr\"umper}, Efremov, \& Sunyaev}]{kretschmar96a}
Kretschmar, P., Pan, H.~C., Kendziorra, E., {et~al.} 1996, A\&A

\bibitem[{Kreykenbohm {et~al.}(2000)Kreykenbohm, Kretschmar, Wilms, Staubert,
  Heindl, Gruber, \& Rothschild}]{kreykenbohm00b}
Kreykenbohm, I., Kretschmar, P., Wilms, J., {et~al.} 2000, in Proc. Conf. X-ray
  Astronomy 99 - Stellar Endpoints, AGN and the Diffuse Background (Gordon \&
  Breach)

\bibitem[{Kreykenbohm {et~al.}(1999)Kreykenbohm, Kretschmar, Wilms, Staubert,
  Kendziorra, Gruber, Heindl, \& Rothschild}]{kreykenbohm99a}
Kreykenbohm, I., Kretschmar, P., Wilms, J., {et~al.} 1999, A\&A, 341, 141

\bibitem[{Leahy {et~al.}(1983)Leahy, Darbo, Elsner, Weisskopf, Sutherland,
  Kahn, \& Grindlay}]{leahy83a}
Leahy, D.~A., Darbo, W., Elsner, R.~F., {et~al.} 1983, ApJ, 266, 160

\bibitem[{Levine {et~al.}(1996)Levine, Bradt, Cui, Jernigan, Morgan, Remillard,
  Shirey, \& Smith}]{levine96a}
Levine, A.~M., Bradt, H., Cui, W., {et~al.} 1996, ApJ, 469, L33

\bibitem[{Maisack {et~al.}(1997)Maisack, Grove, Kendziorra, Kretschmar,
  Staubert, \& Strickman}]{maisack97a}
Maisack, M., Grove, J.~E., Kendziorra, E., {et~al.} 1997, A\&A, 325, 212

\bibitem[{Makishima {et~al.}(1992)Makishima, Mihara, Nagase, \&
  Murakami}]{makishima92a}
Makishima, K., Mihara, T., Nagase, F., \& Murakami, T. 1992, in Frontiers of
  X-Ray Astronomy, Proc, {XXVIII} Yamada Conf., ed. Y.~Tanaka \& K.~Koyama,
  Frontiers Science Series, 2, 23--32

\bibitem[{Makishima {et~al.}(1999)Makishima, Mihara, Nagase, \&
  Tanaka}]{makishima99a}
Makishima, K., Mihara, T., Nagase, F., \& Tanaka, Y. 1999, ApJ, 525, 978

\bibitem[{McClintock {et~al.}(1976)McClintock, Rappaport, Joss, Bradt, Buff,
  Clark, Hearn, Lewin, Matilsky, Mayer, \& Primini}]{mcclintock76a}
McClintock, J.~E., Rappaport, S., Joss, P.~C., {et~al.} 1976, ApJ, 206, L99

\bibitem[{M{\'e}sz{\'a}ros \& Nagel(1985)}]{meszaros85a}
M{\'e}sz{\'a}ros, P. \& Nagel, W. 1985, ApJ, 298, 147

\bibitem[{Mihara(1995)}]{mihara95a}
Mihara, T. 1995, PhD thesis, RIKEN, Tokio

\bibitem[{Mihara {et~al.}(1998)Mihara, Makishima, \& Nagase}]{mihara98a}
Mihara, T., Makishima, K., \& Nagase, F. 1998, in AllSky Monitor Survey
  Conference, RIKEN, 135--140

\bibitem[{Mytrophanov \& Tsygan(1978)}]{mytrophanov78a}
Mytrophanov, I.~G. \& Tsygan, A.~I. 1978, A\&A, 70, 133

\bibitem[{Nagase(1989)}]{nagase89a}
Nagase, F. 1989, PASJ, 41, 1

\bibitem[{Nagase {et~al.}(1983)Nagase, Hayakawa, Makino, Sato, \&
  Makishima}]{nagase83a}
Nagase, F., Hayakawa, S., Makino, F., Sato, N., \& Makishima, K. 1983, PASJ,
  35, 47

\bibitem[{Nagase {et~al.}(1986)Nagase, Hayakawa, Sato, Masai, \&
  Inoue}]{nagase86a}
Nagase, F., Hayakawa, S., Sato, N., Masai, K., \& Inoue, H. 1986, PASJ, 38, 547

\bibitem[{Nagel(1981{\natexlab{a}})}]{nagel81b}
Nagel, W. 1981{\natexlab{a}}, ApJ, 251, 278

\bibitem[{Nagel(1981{\natexlab{b}})}]{nagel81a}
---. 1981{\natexlab{b}}, ApJ, 251, 288

\bibitem[{Ohashi {et~al.}(1984)Ohashi, Inoue, Koyama, Fumiyoshi~Makino,
  Suszuki, Tanaka, Satio~Hayakawa, \& Yamashita}]{ohashi84a}
Ohashi, T., Inoue, H., Koyama, K., {et~al.} 1984, PASJ, 36, 699

\bibitem[{Orlandini {et~al.}(1998)Orlandini, {Dal Fiume}, Frontera, Cusumano,
  {Del Sordo}, Giarrusso, Piraino, Segreto, Guainazzi, \& Piro}]{orlandini97c}
Orlandini, M., {Dal Fiume}, D., Frontera, F., {et~al.} 1998, A\&A, 332, 121

\bibitem[{Rappaport \& McClintock(1975)}]{rappaport75a}
Rappaport, S. \& McClintock, J.~E. 1975, IAU Circ., 2794

\bibitem[{Raubenheimer(1990)}]{raubenheimer90a}
Raubenheimer, B.~C. 1990, A\&A, 234, 172

\bibitem[{Rothschild {et~al.}(1998)Rothschild, Blanco, Gruber, Heindl,
  MacDonald, Marsden, Pelling, Wayne, \& Hink}]{rothschild98a}
Rothschild, R.~E., Blanco, P.~R., Gruber, D.~E., {et~al.} 1998, ApJ, 496, 538

\bibitem[{Santangelo {et~al.}(1999)Santangelo, Segreto, Giarrusso, {Dal Fiume},
  Orlandini, Parmar, Oosterbroek, Bulik, Mihara, Campana, Israel, \&
  Stella}]{santangelo99b}
Santangelo, A., Segreto, A., Giarrusso, S., {et~al.} 1999, ApJ, L85

\bibitem[{Stark(1997)}]{stark97b}
Stark. 1997, {PCABackEst} Information Homepage,
  \url{http://ww2.lafayette.edu/~starkm/pca/pcabackest.html}

\bibitem[{Staubert {et~al.}(1980)Staubert, Kendziorra, Pietsch, C.~Reppin, \&
  Voges}]{staubert80a}
Staubert, R., Kendziorra, E., Pietsch, W., C.~Reppin, J.~T., \& Voges, W. 1980,
  ApJ, 239, 1010

\bibitem[{Stickland {et~al.}(1997)Stickland, Lloyd, \&
  Radziun-Woodham}]{stickland97a}
Stickland, D., Lloyd, C., \& Radziun-Woodham, A. 1997, MNRAS, 286

\bibitem[{Sturner \& Dermer(1994)}]{sturner94a}
Sturner, S.~J. \& Dermer, C.~D. 1994, A\&A, 284, 161

\bibitem[{Sunyaev \& Titarchuk(1980)}]{sunyaev80a}
Sunyaev, R.~A. \& Titarchuk, L.~G. 1980, A\&A, 86, 121

\bibitem[{Tanaka(1986)}]{tanaka86a}
Tanaka, Y. 1986, in Radiation Hydrodynamics in Stars and Compact Objects, ed.
  D.~Mihalas \& K.-H.~A. Winkler, IAU Coll. No.~89 (Springer Verlag), 198

\bibitem[{{van Kerkwijk} {et~al.}(1995){van Kerkwijk}, {van Paradijs},
  Zuiderwijk, Hammerschlag-Hensberge, Kaper, \& Sterken}]{kerkwijk95a}
{van Kerkwijk}, M.~H., {van Paradijs}, J., Zuiderwijk, E.~J., {et~al.} 1995,
  A\&A, 303, 483

\bibitem[{White {et~al.}(1983)White, Swank, \& Holt}]{white83a}
White, N.~E., Swank, J.~H., \& Holt, S.~S. 1983, ApJ, 270, 711

\bibitem[{Willingale {et~al.}(2001)Willingale, Aschenbach, Griffiths, Sembay,
  Warwick, Becker, Abbey, \& Bonnet-Bidaud}]{willingale01a}
Willingale, R., Aschenbach, B., Griffiths, R.~G., {et~al.} 2001, A\&A, 365,
  L212

\bibitem[{Wilms {et~al.}(1999)Wilms, Nowak, Dove, Fender, \& {Di
  Matteo}}]{wilms99b}
Wilms, J., Nowak, M.~A., Dove, J.~B., Fender, R.~P., \& {Di Matteo}, T. 1999,
  ApJ, 522, 460

\end{thebibliography}
\end{document}